\documentclass[twocolumn]{aastex63}

\usepackage{amsmath}
\usepackage{xcolor}
\usepackage{CJK}
\usepackage{tablefootnote}

\shorttitle{White Dwarf Water Fight}
\shortauthors{Trierweiler et al.}
\graphicspath{{}{Figures/}}

\begin{document}
\begin{CJK}{UTF8}{gbsn}
\title{Do White Dwarfs Sample Water-Rich Planetary Material?}

\author[0000-0002-4872-1021]{Isabella L. Trierweiler}
\affiliation{Department of Physics and Astronomy, University of California, Los Angeles, CA 90095, USA}
\affiliation{Department of Astronomy, Yale University, New Haven, CT 06511, USA}
\correspondingauthor{Isabella L. Trierweiler}
\email{isabella.trierweiler@yale.edu}

\author[0000-0001-9834-7579]{Carl Melis}
\affiliation{Department of Astronomy and Astrophysics, University of California, San Diego, La Jolla CA 92093-0424, USA}

\author[0000-0002-3307-1062]{\'{E}rika Le Bourdais}
\affiliation{D\'epartement de Physique, Universit\'e de Montr\'eal, Montr\'eal, QC H3C 3J7, Canada}
\affiliation{Institut Trottier de recherche sur les exoplan\`etes (IREx), Universit\'e de Montr\'eal, Montr\'eal, QC H3C 3J7, Canada}

\author[0000-0003-4609-4500]{Patrick Dufour}
\affiliation{D\'epartement de Physique, Universit\'e de Montr\'eal, Montr\'eal, QC H3C 3J7, Canada}
\affiliation{Institut Trottier de recherche sur les exoplan\`etes (IREx), Universit\'e de Montr\'eal, Montr\'eal, QC H3C 3J7, Canada}

\author[0000-0001-6654-7859]{Alycia J. Weinberger}
\affiliation{Earth and Planets Laboratory, Carnegie Institution for Science, 5241 Broad Branch Rd NW, Washington, DC 20015, USA }

\author[0000-0002-2761-3005]{Boris T. G\"ansicke}
\affiliation{Department of Physics, University of Warwick, Coventry, CV4 7AL, UK}

\author[0000-0002-6428-4378]{Nicola Gentile-Fusillo}
\affiliation{Department of Physics, Universit\`a degli Studi di Trieste, I-34127 Trieste, Italy}



\author[0000-0002-8808-4282]{Siyi Xu (许\CJKfamily{bsmi}偲\CJKfamily{gbsn}艺)}
\affiliation{Gemini Observatory/NSF's NOIRLab, 670 N A'ohoku Place, Hilo, HI
96720, USA}

\author[0000-0003-1748-602X]{Jay Farihi}
\affiliation{Department of Physics \& Astronomy, University College London, Gower Street, London WC1E 6BT, UK}

\author[0000-0001-6515-9854]{Andrew Swan}
\affiliation{Department of Physics, University of Warwick, Coventry CV4 7AL, UK}

\author[0000-0002-7670-670X]{Malena Rice}
\affiliation{Department of Astronomy, Yale University, New Haven, CT 06511, USA}

\author[0000-0002-1299-0801]{Edward D. Young}
\affiliation{Department of Earth, Planetary, and Space Sciences, University of California, Los Angeles, Los Angeles, CA 90095, USA}

\begin{abstract}
Polluted white dwarfs offer a unique way to directly probe the compositions of exoplanetary bodies. We examine the water content of accreted material using the oxygen abundances of 51 highly polluted white dwarfs. Within this sample, we present new abundances for three H-dominated atmosphere white dwarfs that showed promise for accreting water-rich material. Throughout, we explore the impact of the observed phase and lifetime of accretion disks on the inferred elemental abundances of the parent bodies that pollute each white dwarf. Our results indicate that white dwarfs sample a range of dry to water-rich material, with median uncertainties in water mass fractions of $\approx$15\%. Amongst the He-dominated white dwarfs, 35/39 water abundances are consistent with corresponding H abundances. While for any individual white dwarf it may be ambiguous as to whether or not water is present in the accreted parent body, when considered as a population the prevalence of water-rich bodies is statistically robust. The population as a whole has a median water mass fraction of $\approx$25\%, and enforcing chondritic parent body compositions, we find that 31/51 WDs are likely to have non-zero water concentrations. This conclusion is different from a similar previous analysis of white dwarf pollution and we discuss reasons why this might be the case. Pollution in H-dominated white dwarfs continues to be more water-poor than in their He-dominated cousins, although the sample size of H-dominated white dwarfs remains small and the two samples still suffer a disjunction in the range of host star temperatures being probed.
\end{abstract}


\section{Introduction}\label{section:intro}

The growing population of super-Earths and sub-Neptunes, and advancements in connecting models of planetary interiors to mass/radius estimates,  has sparked an ongoing discussion on the prevalence of water worlds \citep[e.g.][]{Unterborn2018, Kempton2023, Piette2023, Boldog2024, Chakrabarty2024}. However, degeneracies in modeling bulk densities often make it difficult to distinguish between water-rich planets and planets with light elements stored in their interiors \citep[e.g.][]{ Dorn2021, Schlichting2022}. 

The relative abundances of rock-forming elements found in the atmospheres of white dwarfs (WDs) that have been polluted by accretion of surrounding planetary material offer an independent method for studying exoplanetary water abundances in extrasolar rocky bodies (e.g., asteroids, comets, or moons). The abundance of water in the parent bodies accreted by WDs is typically inferred from excess abundances of O relative to other rock-forming elements \citep{Klein2010}, and a number of WDs have been identified as having accreted icy or watery bodies based on oxygen excesses \citep[e.g.][]{Farihi2013, Raddi2015, Xu2017, Hoskin2020, Klein2021, Hollands2022}. Observations of hydrogen persistent in the atmospheres of helium-dominated WDs have also been used to assess water abundances. However, while observations of H are fairly common in He atmospheres, accretion rates averaged over the cooling age of the WD are frequently so low that the associated water accrued by accretion would be negligible in comparison to the accretion rates of rocky material \citep{Jura2012b, GentileFusillo2017}. 
 
In this work we leverage a large sample of polluted WDs with O abundances to gain a better understanding of the distribution of water abundances amongst WD polluters. Accurately recovering parent body compositions from polluted WDs relies on assumptions made about the accretion and settling rates in the WD atmosphere \citep[e.g.][]{Doyle2020,obrien25}; we therefore also test how different phases of the accretion process influence inferred water abundances. We test accretion phases by assigning intervals of increasing mass with time (mass-buildup phase), decreasing mass with time (mass-settling phase), and steady state accretion to each WD and comparing the effects of these processes on inferred parent body compositions and implied water contents.

We quantify our comparison using a Markov-chain Monte Carlo (MCMC) sampling to outline the posterior distributions of characteristic parameters in our accretion model and to thereby find the best fit to the observations. In the MCMC tests, we find that if we assume most WD pollution comes from parent bodies with chondritic compositions, then most WDs are in or near a steady state phase, though the distributions are broad.

We interpret our resulting water abundance distributions as evidence
for significant O excesses that result in a wide range of water mass fractions, with a peak in the overall distribution at around $20-30\%$.
This is in contrast to what was concluded by \citet{Brouwers2023} following a
similar analysis, although the distributions we compute are not incompatible
with those presented in \citet{Brouwers2023}. Potential
origins for the differing conclusions on water abundances 
are discussed in Section \ref{section:evaporation}.

This paper is structured as follows. Section \ref{section:obs} presents new abundances for three H-dominated WDs where preliminary analyses suggested
they could potentially be accreting water-rich planetary material; one such object
is confirmed to indeed be water-rich. 
In Section \ref{section:methods} we collect all published WDs with O abundances and develop a model to assess the impact of accretion and 
settling on calculating water abundances.
We report water mass fractions for all studied objects in Section \ref{section:results}, discuss our water results and our model's implications for WD accretion in Section \ref{section:discussion}, and summarize these findings in Section \ref{section:conclusion}.

\section{Abundances for three H-dominated White Dwarfs}\label{section:obs}

In this section we present new observations for three WD stars.
These objects were part of an effort to identify more H-dominated
atmosphere WDs (hereafter H-WDs) with evidence for water pollution as there seemed
to be a dearth of such systems in the literature (counterintuitive when
one considers that H-WDs are supposed to be the dominant variety; e.g., \citealt{Koester2015}).

\subsection{Observations and Data Reduction}\label{subsection:obs}

Table \ref{tab:newHWDs} provides the observation details for each of the WDs. Instruments used include HIRES on the Keck I Telescope at Maunakea Observatory \citep{Vogt1994} for optical observations of two stars, X-Shooter on the Very Large Telescope \citep{dodorico06, Vernet2011} for one star, 
the MIKE spectrograph on the Clay Telescope at Las Campanas Observatory \citep{Bernstein2003} for one star, and the 
{\it Hubble Space Telescope} ({\it HST}) Cosmic Origins Spectrograph (COS; \citealt{green12}) for far-ultraviolet coverage of two stars.

\centerwidetable
\begin{deluxetable*}{l|l|l|l|c|r}
\tablecaption{WD Observation Data}\label{tab:newHWDs} 
\tablehead{\colhead{WD Name} & \colhead{Date} & \colhead{Instrument} & \colhead{Coverage (\r{A})} & \colhead{Resolving Power} & \colhead{S/N\textsuperscript{a}}} 
\startdata
  WD\,0145+234 & 2019/09/21 & HIRES (blue) & 3130 - 5945 & 37000 & 44 \\
    & 2019/12/09  & HIRES (red) & 4715 - 8985 & 37000 & 140 \\ 
   &  2019/11/27 & COS (G130M) & 1130 - 1430 & $\sim$15000 & 17 \\ 
    & 2019/11/28 & COS (G160M) &  1410 - 1800 & $\sim$15000 & 24 \\ 
\hline
   WD\,0842+572 & 2019/12/09  & HIRES (red) & 4250 - 5990 & 37000 & 60 \\ 
   & 2021/12/13 & HIRES (blue) & 3110 - 5950 & 37000 & 26 \\
\hline
 WD\,J0649$-$7624 & 2019/01/12 & X-Shooter (UV) & 3000 - 5500 & $\sim$5500 & 50 \\
     & 2019/01/12 & X-Shooter (Vis) & 5300 - 10000 & $\sim$9000 & 70\\ 
   & 2019/03/21 & MIKE & 3350 - 8850 & $\sim$25000 & 25 \\
    & 2019/12/25 & COS (G130) & 1130 - 1430 & $\sim$15000 & 9 \\
\enddata
\tablenotetext{a}{Signal-to-noise ratio per pixel (S/N) measured at  1350 \r{A} (COS G130M), 1500 \r{A} (COS G160M), 3500 \r{A} (HIRES blue, X-Shooter UV),  5000 \r{A} (HIRES red, MIKE), and 6300 \r{A} (X-Shooter Vis).}
\end{deluxetable*}

\begin{figure*}
    \centering
    \includegraphics[trim={1.3cm 5.9cm 1.1cm 6.6cm},clip,width=1\linewidth]{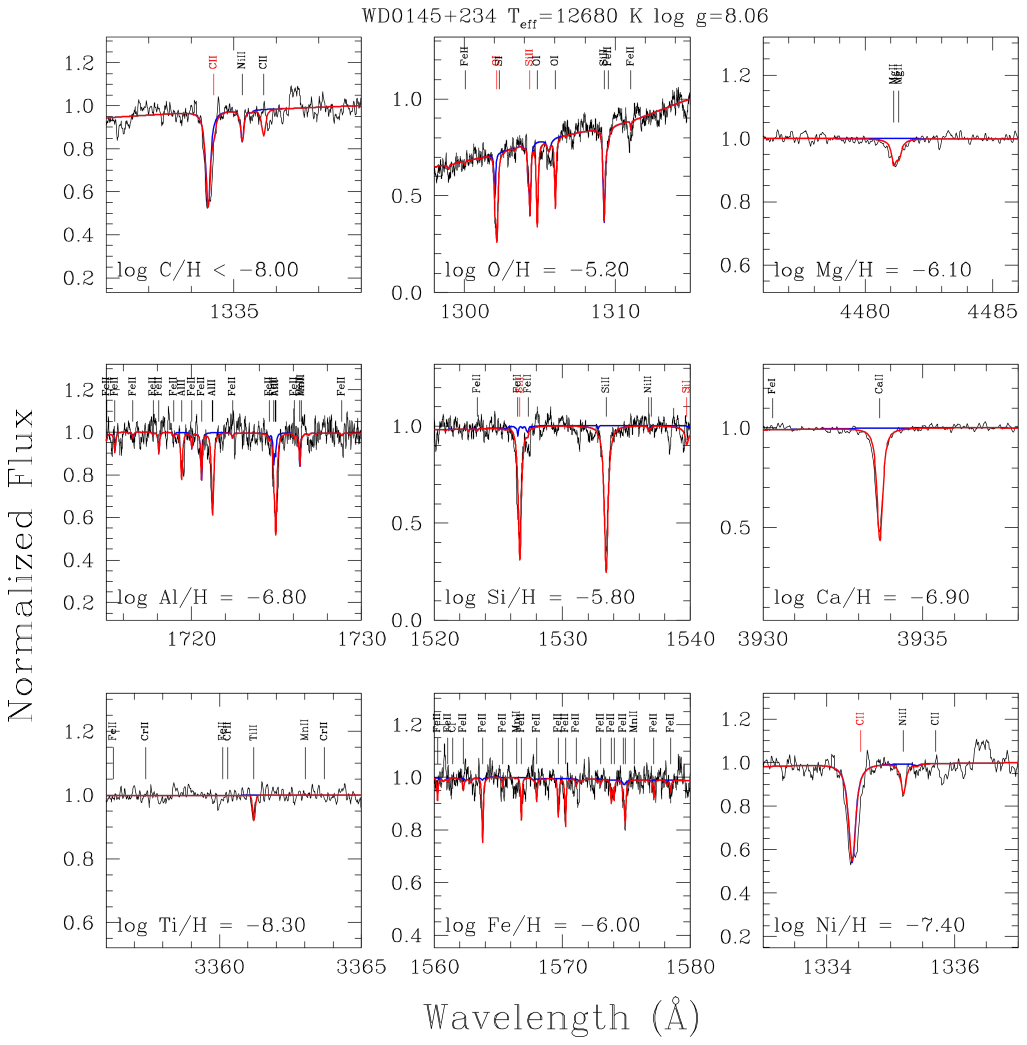}
    \caption{A selection of fits for the absorption lines for WD\,0145+234. The red line shows the modeled spectrum, while the blue line shows the fit with the labeled element (listed as $\log \rm (X/H)$ in each panel) removed. Species identified in red indicate ground-state transitions; these may be due to the ISM.}\label{WD0145_optical}
\end{figure*}

\begin{figure*}
    \centering
    \includegraphics[trim={1.3cm 5.9cm 1.1cm 6.5cm},clip,width=1\linewidth]{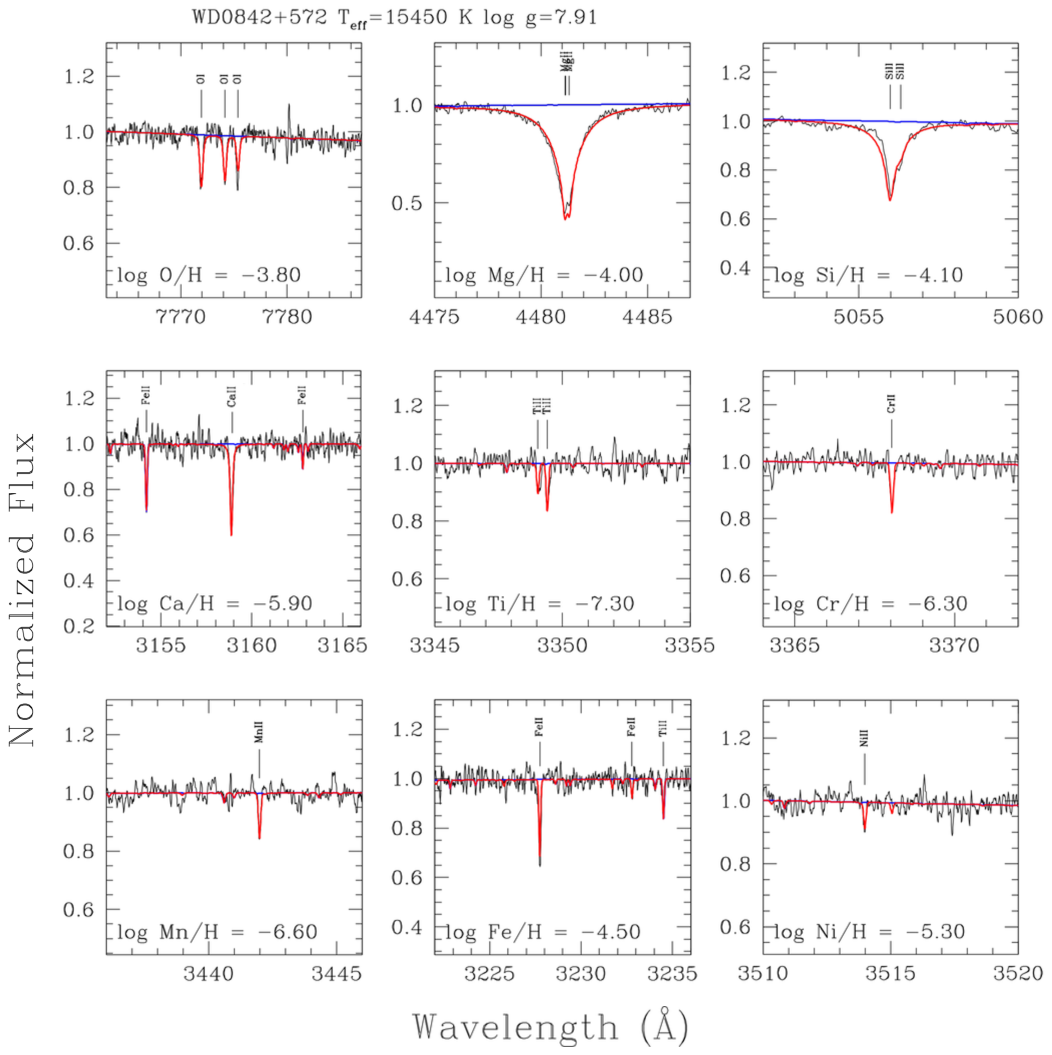}
    \caption{A selection of fits for the absorption lines in WD\,0842+572. Colored lines and labels are as described in the caption of Figure \ref{WD0145_optical}.}\label{WD0842}
\end{figure*}

\begin{figure*}
    \centering
    \includegraphics[trim={0.9cm 11.6cm 1.1cm 6.6cm},clip,width=1\linewidth]{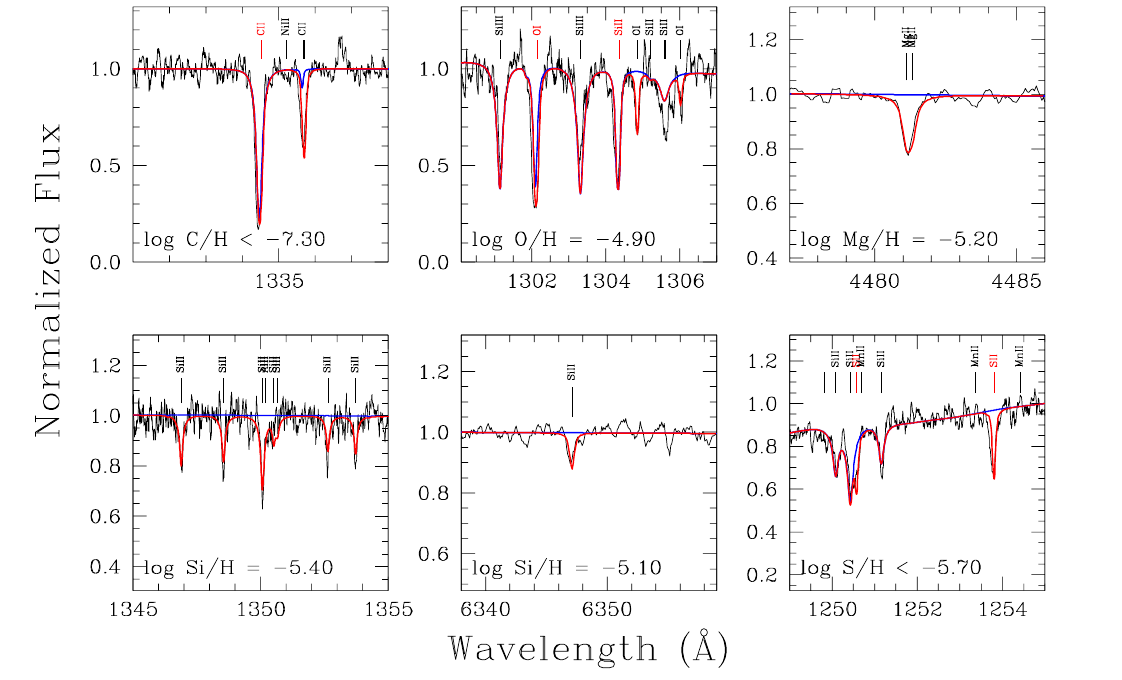}
    \caption{A selection of fits for the absorption lines for WD\,J0649$-$7624
    for the case of the spectroscopic solution. Colored lines and labels are as described in the caption of Figure \ref{WD0145_optical}. See Section \ref{section:j0649} for a discussion of interstellar medium absorption and how limits are assessed for C and S.}\label{J0649}
\end{figure*}

HIRES raw data are reduced using the
{\sf MAKEE} software package which outputs spectra in vacuum wavelengths that have been shifted to the heliocentric reference frame. While HIRES wavelength coverage is quoted as continuous in Table \ref{tab:newHWDs}, there are gaps in coverage between each of the three CCDs and sometimes between red orders. Further processing in {\sf IRAF} as described in \citet{Klein2021} and references therein removes instrumental profiles in the spectra, bringing overlapping order segments into agreement, then combines all orders of every
exposure to generate a final spectrum for analysis.

X-shooter data are reduced using the standard procedures in Reflex \citep{ESOReflex}\footnote{\dataset[https://www.eso.org/sci/software/esoreflex/]{https://www.eso.org/sci/software/esoreflex/}} version 2.9.1, with X-shooter pipeline version 3.2 and molecfit version 1.1.0 to remove telluric lines \citep{Smette2015, Kausch2015}. Standard X-shooter data reduction techniques are employed with default settings to extract and wavelength calibrate each spectrum.
Relative flux calibration on the science spectrum is performed
with the use of the spectrophotometric
standard EG\,21 / CPD-69 177 to derive the instrumental response function. Although
X-shooter coverage extends to the thermal-infrared, data beyond
$\approx$1\,$\mu$m are unusable due to low recorded signal.

MIKE data are reduced using the MIKE pipeline by D.\ Kelson \citep{Kelson2000, Kelson2003} which provides CCD processing, flat fielding with nightly flat field images, extraction, wavelength calibration with a contemporaneous ThAr lamp frame, and blaze correction. Three individual spectral images are combined before extraction. A final order-merged spectrum is obtained after normalizing orders by fitting a polynomial to the white dwarf continuum and averaging data in
overlap regions.

COS data were processed with the {\sf \verb+CALCOS+} pipeline 3.3.7, coadded with the use of the {\sf IDL} script {\sf \verb+COADD_X1D+} \citep{Danforth2010,Keeney2012}, and 
then smoothed with a boxcar of six pixels to produce a final spectrum that is flux calibrated, in vacuum wavelengths, and corrected to the heliocentric reference frame. Following \citet{Jura2012}, we use the {\sf timefilter} module to extract the night-time portion of the data around the O~I lines between 1300 and 1308\,\AA\ to help mitigate terrestrial day airglow emission. This is mostly done as a sanity check on the abundance 
for the one uncontaminated O~I line near 1152\ \AA\ which otherwise would be the only line to inform the ultraviolet oxygen abundances. The specific observations analyzed can be accessed via \dataset[doi: 10.17909/2cqw-zs48]{https://doi.org/10.17909/2cqw-zs48}.

\centerwidetable
\begin{deluxetable*}{l|r|r|r|r|c|c}
\tablecaption{WD Parameters}\label{tab:newwd_param}
\tablehead{\colhead{WD Name} & \colhead{R.A. (J2000)} & \colhead{Dec. (J2000)} & \colhead{$T_\mathrm{{eff}}$ (K)} & \colhead{$\log g$ (cgs)} & \colhead{$M_\mathrm{{WD}}$ ($M_\mathrm{\odot}$)} & \colhead{$\log$ $M_\mathrm{{env}/M_{WD}}$ } } 
\startdata
WD\,0145+234 & 01 47 54.82 & +01 45 08.04 & 12680 $\pm$ 80 & 8.06 $\pm$ 0.01 & 0.64 & $-$15.706 \\
WD\,0842+572 & 08 46 02.40 & +57 03 28.30 & 15400 $\pm$ 70 & 7.91 $\pm$ 0.01 & 0.56 & $-$16.379\\
WD\,J0649$-$7624 (phot) & 06 49 34.85 & $-$76 24 58.10 & 21908 $\pm$ 483& 8.09 $\pm$ 0.02 & 0.68 & $-$16.526 \\
WD\,J0649$-$7624 (spec) & 06 49 34.85 & $-$76 24 58.10 & 23504 $\pm$ 690 & 8.17 $\pm$ 0.10 & 0.73 & $-$16.583 \\
\enddata
\tablenotetext{}{$T_\mathrm{{eff}}$ and $\log g$ are calculated from photometric fits for each WD. $M_\mathrm{{WD}}$ and the convective zone mass ratio ($\log M_\mathrm{{env}/M_{WD}}$) are drawn from the Montreal White Dwarf Database \citep{Dufour2017}. Both the photometric and spectroscopic solutions are reported for WD\,J0649$-$7624 as discussed in Section \ref{section:j0649}.}
\end{deluxetable*}

We determine stellar parameters $T_{\mathrm{eff}}$ and log $g$ by fitting Gaia DR3 parallaxes \citep{GaiaCollaboration2023} and collections of photometry data for each star, including PanSTARRS \citep{Flewelling2020}, SDSS \citep{Alam2015}, and SkyMapper \citep{Onken2024} observations. The best-fit parameters for each WD are listed in Table \ref{tab:newwd_param}. Calculated stellar parameters can vary based on whether photometric or spectroscopic data are used \citep[e.g.,][]{Tremblay2019, Cukanovaite2021}; we select the photometric solutions as these have typically provided more consistent and reliable results, particularly when using SDSS and PanSTARRS data \citep{Genest-Beaulieu2019, Izquierdo2023}.
For the case of WD\,J0649$-$7624 we consider both the photometric and spectroscopic
solutions as discussed in detail below.

Using the fitted stellar parameters, we then obtain elemental abundances by iteratively fitting the synthetic spectra to the observed data. 
Measured radial velocities and equivalent widths (EW) for the full list of detected lines suspected to originate from the white dwarf photosphere are given in Appendix \ref{appsection:line_lists}. See \citet{Dufour2007,Dufour2010} and \citet{Coutu2019} for detailed discussion of the fitting procedures. Lines having different radial velocities than that of the white dwarf are non-photospheric in origin and are often attributed to the interstellar medium (ISM). It has been shown that they can impact the obtained photospheric abundances when these lines are blended with photospheric lines \citep{LeBourdais2024}. To limit the impact on our obtained abundances, we used a Voigt profile to reproduce the ISM lines before fitting the photospheric lines.
We show a sample of fitted spectral lines in Figures \ref{WD0145_optical},
\ref{WD0842}, and \ref{J0649}. 

Table \ref{tab:newHWds_abundance} lists the log abundance ratios relative to H for each observed element, averaged across all observed spectral lines. Standard deviations for the range of abundances determined for the spectral lines for each element typically range from about 0.02 to 0.15 dex, and we conservatively apply uncertainties of 0.2 dex to the abundances in this work.  For elements with lines in both the UV and optical, we use the mean value for the abundance and water analysis in this work.
For Si in WD\,J0649$-$7624 we find a significant discrepancy in abundances between the visible and UV fits, about 0.6 dex.
Such discrepancies for Si have been found before; see e.g., \citet{Xu2019} and \citet{Rogers2024a}.
We adopt the average of the two Si values and an uncertainty that encompasses
both UV and optical values throughout the rest of this work.   

\centerwidetable
\begin{deluxetable*}{l|r|r|r|r|r|r|r|r|r|r|r}
\tablecaption{Observed atmospheric elemental abundances in $\log \left(\mathrm{n_Z/n_H}\right)$ and respective settling timescales\textsuperscript{a}}\label{tab:newHWds_abundance}
\tablehead{\colhead{} & \multicolumn{3}{l}{WD\,0145+234} & \multicolumn{2}{l}{WD\,0842+572} & \multicolumn{3}{l}{WD\,J0649$-$7624 phot} & \multicolumn{3}{l}{WD\,J0649$-$7624 spec} \\ 
\colhead{Z} & \colhead{Opt} & \colhead{UV} & \colhead{$\log \mathrm{\tau_{Z}}$(yr)} & \colhead{Opt} & \colhead{$\log \mathrm{\tau_{Z}}$(yr)} & \colhead{Opt} & \colhead{UV} & \colhead{$\log \mathrm{\tau_{Z}}$(yr)} & \colhead{Opt} & \colhead{UV} & \colhead{$\log \mathrm{\tau_{Z}}$(yr)} } 
\startdata
C & $-$ & $<$-8.0 & -1.549 & $-$ & -2.050 & $-$ & $<$-7.2 & -2.181 & $-$ & $<$-7.3 & -2.197 \\
N & $-$ & $<$-7.0 & -1.678 & $-$ & -2.133 & $-$ & $<$-5.3 & -2.331 & $-$ & $<$-5.4 & -2.373 \\
O & $-$ & -5.20 & -1.767 & -3.8 & -2.217 & $<$-5.0 & -5.0 & -2.429 & $<$-4.0 & -4.9 & -2.507 \\
Mg & -6.1 & $-$ & -1.500 & -4.0 & -1.905 & -3.7    & $-$  & -2.122 & -5.2 & $-$ & -2.220 \\
Al & $-$ & -6.8 & -1.522 & $<$-5.5 & -2.006 & $<$-4.0 & $<$-6.7 & -2.142 & $<$-4.5 & $<$-6.8 & -2.205 \\
Si & $-$ & -5.8 & -1.578 & -4.1 & -2.111 & -5.2 & -5.8 & -2.197 & -5.1 & -5.7 & -2.269 \\
P & $-$ & $<$-7.0 & -1.629 & $-$ & -2.244 & $-$ & $<$-7.4 & -2.253 & $-$ & $<$-7.3 & -2.331 \\
S & $-$ & $<$-7.0 & -1.686 & $-$ & -2.377 & $-$ & $<$-5.0 & -2.296 & $-$ & $<$-5.0 & -2.363 \\
Ca & -6.9 & -7.1 & -1.686 & -5.9 & -2.091 & $<$-5.0 & $-$ & -2.337 & $<$-6.0 & $-$ & -2.436 \\
Ti & -8.3 & $-$ & -1.745 & -7.3 & -2.173 & $<$-4.0 & $-$ & -2.373 & $<$-5.3 & $-$ & -2.431 \\
Cr & $<$-7.2 & $-$ & -1.787 & -6.3 & -2.200 & $-$ & $-$ & -2.430 & $-$ & $-$ & -2.502 \\
Mn & $<$-7.5 & $-$ & -1.823 & -6.6 & -2.277 & $<$-4.0 & $<$-5.5 & -2.460 & $<$-5.5 & $<$-5.0 & -2.528 \\
Fe & $-$ & -6.0 & -1.850 & -4.5 & -2.362 & $<$-3.5 & $<$-5.5 & -2.480 & $<$-3.5 & $<$-5.5 & -2.548 \\
Ni & $-$ & -7.4 & -1.861 & -5.3 & -2.316 & $<$-3.5 & $<$-7.0 & -2.504 & $<$-4.5 & $<$-6.8 & -2.596 \\
\enddata
\tablenotetext{a}{Settling timescales $\mathrm{\tau_{Z}}$ are drawn from the Montreal White Dwarf Database, using the parameters listed in Table \ref{tab:newwd_param}. Abundances are reported for WD\,J0649$-$7624 for both its photometric and spectroscopic solutions as discussed in Section \ref{section:j0649}.}
\end{deluxetable*}

\begin{figure*}
    \centering
    \includegraphics[width=1\linewidth]{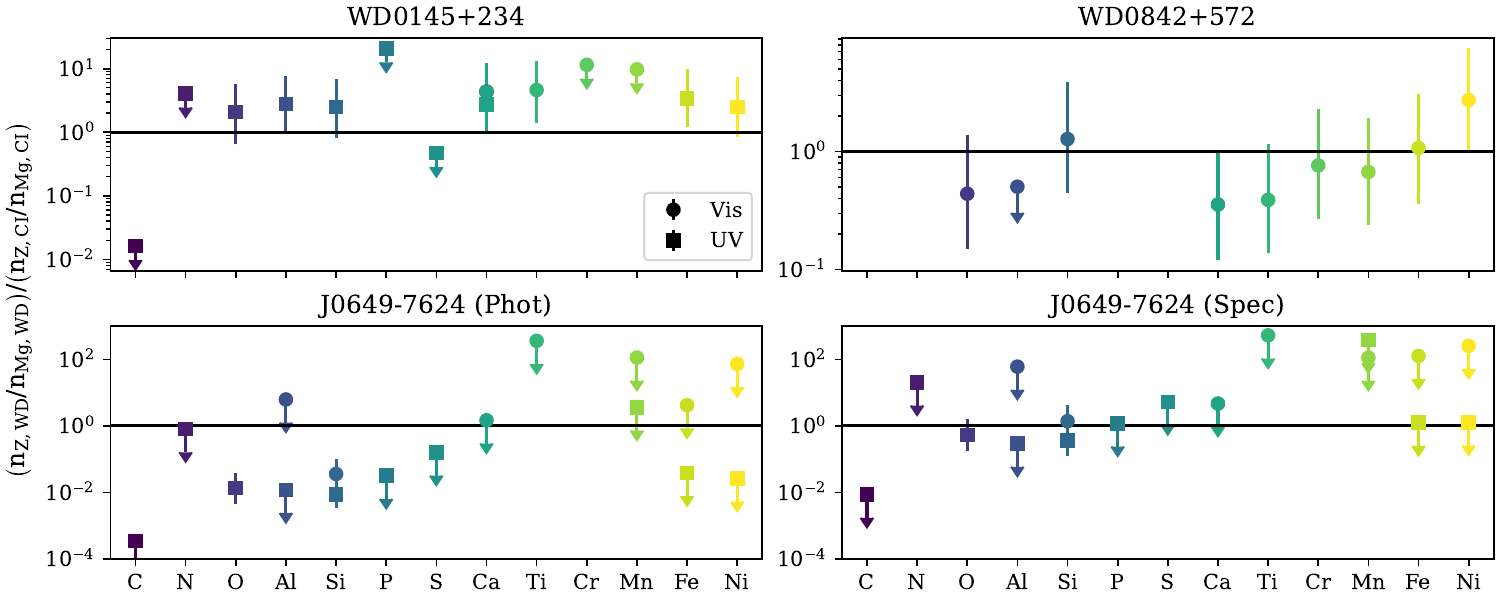}
    \caption{Abundances of the observed H-WDs described in Section \ref{section:obs} assuming the steady state accretion phase, ratioed to Mg and reported relative to CI chondrites with $2\sigma$ uncertainties. The circular points show values determined from optical spectra while the squares show the UV results. All points are colored according to the corresponding element. For clarity, we show only the detection for WD\,J0649$-$7624's O abundance in the photometric and spectroscopic solutions.}\label{zplot_new}
\end{figure*}

\subsection{Discussion of Individual White Dwarfs}

As illustrated in Table \ref{tab:newHWds_abundance}, the settling times for the
three H-WDs are very short ($\lesssim$12~days for WD0145+234
and $\lesssim$5~days for the other two). Given such short settling times, it is
unlikely that these systems are in any accretion phase other than the steady
state. Indeed, despite observation epochs for each object typically spanning at least one and usually numerous settling times, abundances from each data set are generally in agreement. We thus assume these objects to be accreting in the steady state for all further analysis. The following subsections provide detailed discussion on the 
history, abundance patterns, and implications for inferred water abundances for
each of the three WDs.

\begin{figure}
    \centering    \includegraphics[width=1\linewidth]{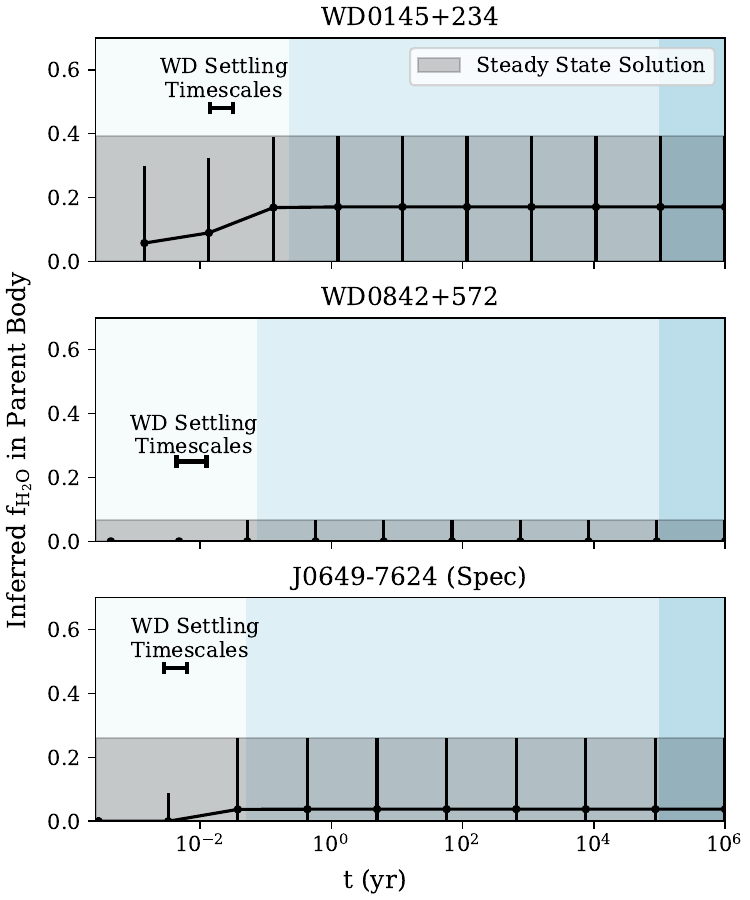}
    \caption{Water mass fractions ($f_{\mathrm{H_2O}}$) calculated from O excesses for the H-WDs described in Section \ref{section:obs}, according to the approach outlined in Section \ref{section:methods}. The ranges of settling timescales for the observed elements on each WD are labeled, and the resulting ranges of buildup, steady state, and settling phases are shown from left to right as the shaded regions. }\label{water_time_new}
\end{figure}

\subsection{WD\,0145+234}
WD\,0145+234 is notable as one of the brightest, nearest, and coolest WDs with a gas disk \citep{Melis2020}. The WD experienced an infrared outburst in 2018, followed by a decline in 2019 \citep{Wang2019}. No changes in the optical photometry of the WD were observed during the outburst. The outburst is therefore attributed to a collision event that replenished the gas and dust in the debris disk, and subsequent stochastic brightening events indicate further collisions within the disk as it returns to a quiescent state \citep{Swan2021}. 

We detect seven elements in the atmosphere of WD\,0145+234 (O, Al, Si, Ca, Ti, Fe, and Ni; Figure \ref{WD0145_optical}).
Considering the system in the steady state accretion phase we find
an accretion rate just under 10$^8$\,g\,s$^{-1}$ and
what appears to be a deficiency in Mg (Mg/Si$\approx$0.42 and Mg/Fe$\approx$0.35 by number, compared to $\approx1$ for bulk Earth), significantly depleted C (C/[Si or Mg or Fe] below the bulk Earth value of $\approx0.01$),
and O between CI Chondrite and the Bulk Earth (O/Si$\approx$6.2 and O/Fe$\approx$5.2 by number). Within the uncertainties, most elements other than C are reasonably in agreement with chondritic compositions (Figure \ref{zplot_new}).
As described in more detail in Section \ref{section:methods}, an O-budget analysis
determines that WD0145+234 has an O excess with median water mass fraction
of $\approx$20\%. This water abundance would support the history of aqueous alteration suggested by the presence of the tentative carbonate feature identified by \cite{Swan2024}.

\subsection{WD\,0842+572}
WD\,0842+572 has a debris disk indicated by an infrared excess \citep{Swan2020}. \cite{Melis2020} find gas emission lines of Si, Mg, Ca, and Fe, consistent with a disk formed by the breakup of a rocky body. They note narrow line profiles and a mix of both neutral and semi-forbidden transition lines, consistent with viewing the disk nearly face-on. There may be some variability in the emission of WD\,0842+572, though the results are not conclusive. 

We detect absorption lines from eight elements (O, Si, Ca, Ti, Cr, Mn, Fe, and Ni) in WD\,0842+572 (Figure \ref{WD0842}). 
While we see a feature consistent with Al~II, the line depth is not significantly larger than the surrounding noise and we therefore report the Al abundance as an upper limit. Al has a minor contribution to the O budget, so considering the Al as a detection instead of upper limit does not significantly change the water abundance.

Assessing WD\,0842+572 in the steady state accretion phase we find an enormous
accretion rate of 2$\times$10$^9$\,g\,s$^{-1}$. The relative abundances of O and Mg are O/[Mg or Si or Fe]$\approx$3 by number and
Mg/Si $\sim$ Mg/Fe $\sim$ 1, matching those of bulk Earth. There appears to be a slight Si enhancement, which is joined by unusual apparent deficits in refractories like Al, Ca, and Ti. Cr and Mn appear roughly consistent with Bulk Earth proportions while Ni appears highly enhanced (about 2.6$\times$ Bulk Earth).
Given the uncertainties, most elements are within $2\sigma$ of chondritic
(Figure \ref{zplot_new}).

An O-budget analysis (see Section \ref{section:methods}) determines that in the steady state WD\,0842+572
presents an O deficit. This would be consistent with a dry,
Earth-like body. Given the bright disk with strong gas emission features
and substantial accretion rate for a H-WD, it would
seem plausible that WD\,0842+572 accretes the remains of a massive rocky body.
This makes one wonder if the body was differentiated and if some significant
fraction of elements like Fe, Ni, Cr, and Mn might be in metal form instead
of rocky oxides. By allowing these elements to be allocated to a metal core-like
component it is possible to drive the O deficit toward being balanced, although
some small deficit still remains even if all Fe is taken to be in metal form.
It is thus very unlikely that WD\,0842+572 is water-rich, and the analysis
of Section \ref{section:methods} suggests a water abundance limit of $<$7\%
by mass.

\subsection{WD\,J0649$-$7624}
\label{section:j0649}
WD\,J0649$-$7624 was first identified as a WD in Gaia DR2 \citep{Jimenez-Esteban2018, GentileFusillo2019}, and was later found to have an infrared excess \citep{Xu2020}.
Based on the identification of infrared excess emission we selected WD\,J0649$-$7624
for observation as part of our large polluted WD survey (\citealt{Melis2018}; \citealt{Doyle2023}), using time allocated with
the Gemini-South GMOS-S spectrograph. GMOS-S data suggested the presence of Mg
(this spectrum can be retrieved from the Montreal White Dwarf Database) and led to further follow-up
observations as reported herein (see Section \ref{subsection:obs}).

For this WD we report detections of O, Mg, and Si with upper limits for all other elements (Figure \ref{J0649}). A number of lines appear to have a significant interstellar medium (ISM) contribution, particularly C and S in the UV (Figure \ref{J0649}). We lack sufficient separation in radial velocity to distinguish between ISM and photospheric lines, however both features are ground-state transitions, and in both cases the ISM contribution can be tuned such that the photospheric contribution to the line profile becomes negligible. 
We use C~III features around 1175\,\AA\ as a cross-check of any suggested
C abundance from modeling the features near 1330\,\AA ; for the temperatures
of WD\,J0649$-$7624 reported in Table \ref{tab:newwd_param} such lines should be
visible if photospheric C is present.
We report upper limits for C and S based on spectral fits that attempt to take into account ISM absorption and the C~III transitions.

Importantly, and in contrast to the other stars
presented above, 
the elemental abundance ratios of WD\,J0649$-$7624 are strongly dependent on whether
the photometric or spectroscopic atmospheric solution is selected (Table \ref{tab:newHWds_abundance}). While the photometric solution is generally preferred
(see above), in the case of WD\,J0649$-$7624 it leads to nonsensical parent body
compositions in the steady state accretion phase with Mg dominating the mass
budget by upwards of 50\%. 
After comparing the photometric and spectroscopic solutions to available
photometry and the flux-calibrated COS spectrum, we conclude that there is unaccounted for reddening toward WD\,J0649$-$7624 that confuses the
photometric solution.
We therefore discard the photometric solution and focus instead 
on the spectroscopic solution results for the remainder of this analysis.

For the spectroscopic solution, WD\,J0649$-$7624 when assessed in the steady state presents an accretion rate on the order of 10$^8$\,g\,s$^{-1}$ and reasonably chondritic or possibly Bulk Earth-like abundance patterns for its polluting parent body (Figure \ref{zplot_new}).
The parent body is C-depleted (C/[Si or Mg]$\lesssim$0.01 by number) and Si appears slightly deficient relative to Mg compared to Bulk Earth. 
The Fe limit allows chondritic or Bulk Earth-like compositions, but otherwise leaves the nature of the parent body unconstrained. For example, it is possible that WD\,J0649$-$7624 could be reminiscent of the Bulk Earth with a slight Mg enhancement and no significant water. Alternatively, Fe could be under-abundant leading to a parent body resembling that found in the SDSS\,J1043+0855 system \citep{Melis2017} and an O excess and hence water content comparable to WD\,0145+234.
Observations in the near-UV could provide the needed Fe detection to robustly assess the parent body composition and water content. In our analysis we
assume a chondritic Fe abundance.

\section{Methods}\label{section:methods}
\subsection{White Dwarf Sample}
\label{sec:3.1}
Our sample consists of 48 published WDs that include the elements O and Mg (for uniform normalization of abundance ratios);
we exclude SDSS\,J0956+5912 \citep{Hollands2022,Swan2023} as it is most likely
deep in the mass-settling phase,
GD\,133 \citep{Xu2014} as Mg is a marginal detection and the Fe non-detection is not restrictive, and some stars from \citet{Izquierdo2023} that are missing O or Mg. We additionally require abundances for at least one other element out of Si, Ca, and Al for the purposes of comparing lithophile element ratios to those in CI chondrites. Our sample also includes the newly acquired abundances for the H-WDs described in Section \ref{section:obs}. 

The references and stellar parameters for the WDs are listed in Table \ref{table:WDsample}. For consistency across the sample set, we draw settling timescales, WD masses, and convective zone mass ratios from the Montreal White Dwarf Database \citep{Dufour2017} based on the effective temperature and log gravity reported by the references in Table \ref{table:WDsample}. 
Many T$_{\rm eff}$ and $\log g$ values would benefit from being revisited with
modern modeling techniques that incorporate {\it Gaia} parallax information and
updated atmospheric structures that take into account metals and H in the case
of He-dominated atmospheres (hereafter He-WDs).
Additionally, there is inherent uncertainty associated with settling times
(e.g., compare \citealt{Koester2020} and \citealt{Dufour2017}).

Elemental abundances for polluted WDs are typically reported in the literature as the logs of abundances by number ratio of metal Z to atmospheric H or He,  $\log \mathrm{\left(  Z/H(e) \right)}$ \citep[e.g.,][]{Jura2014}. Throughout this work, we propagate uncertainties in the log elemental abundances to simple ratios and eventually water abundances by taking 1000 random draws of the log elemental abundances assuming a normal distribution with the reported abundance and uncertainty as the mean and standard deviation. After converting the random draws to the quantity of interest, we report the median, lower, and upper uncertainties of the new quantity as the 50, 16, and 84th percentiles of the converted random draws, respectively.

Throughout this work we assume that the material being accreted by each WD is made up of a single body, or that the mass observed in the atmosphere is primarily due to a single accretion event. While exceptions may be possible (e.g., \citealt{Johnson2022}), the results of \cite{Trierweiler2022} generally support this assumption and show that even when accretion is driven by many smaller objects, such as during the accretion of a debris belt, we are most likely observing the WD soon after the latest accretion event when a single object still dominates the observable material. 

\startlongtable
\tabletypesize{\scriptsize}
\centerwidetable
\begin{deluxetable*}{l|l|l|l|l|l|l|l}
\tablecaption{WD sample; parameters are collected from the references listed in the table. Throughout this work we group WDs by the dominant element in their atmospheres (H or He), and we list them here in order of increasing $T_\mathrm{{eff}}$. WDJ names include the RA and Dec (J2000) for each object, and are formatted as JHHMMSS.ss+DDMMSS.ss \citep{GentileFusillo2021}.}\label{table:WDsample} 
\tablehead{\colhead{Name in Reference} & \colhead{Gaia DR3 Source ID} & \colhead{ 
WDJ name} & \colhead{Reference} & \colhead{Type} & \colhead{$T_\mathrm{{eff}}$ (K)} & \colhead{log g} & \colhead{log H/He}}
\startdata
G29$-$38&2660358032257156736&WD\,J232847.64+051454.24&\citealt{Xu2014}&H&11820&8.15 &  \\
WD\,0145+234&291057843317534464&WD\,J014754.82+233943.60&This Work&H&12680&8.06 & \\
WD\,0842+572&1037258898615762048&WD\,J084602.47+570328.64&This Work&H&15400&7.91 & \\
J0611$-$6931&5279484614703730944&WD\,J061131.70$-$693102.15&\citealt{Rogers2024}&H&16530&7.81 & \\
SDSS\,J1043+0855&3869060540584643328&WD\,J104341.53+085558.35&\citealt{Melis2017}&H&18330&8.05 & \\
PG\,1015+161&3888723386196630784&WD\,J101803.83+155158.57&\citealt{Gansicke2012}&H&19226&8.04 & \\
J0510+2315&3415788525598117248&WD\,J051002.15+231541.42&\citealt{Rogers2024}&H&20130&8.13 & \\
SDSS\,1228+1040&3904415787947492096&WD\,J122859.93+104033.04&\citealt{Gansicke2012}&H&20900&8.15 & \\
GALEX\,1931+0117&4287654959563143168&WD\,J193156.93+011744.11&\citealt{Gansicke2012}, &H&21200&7.91 & \\
& & & \citealt{Melis2011} & & & & \\
J0006+2858&2860923998433585664&WD\,J000634.71+285846.54&\citealt{Rogers2024}&H&22840&7.86  & \\
PG\,0843+516&1029081452683108480&WD\,J084702.29+512853.35&\citealt{Gansicke2012}&H&23095&8.17 & \\
WD\,J0649$-$7624&5212764251961409664&WD\,J064935.06$-$762500.02&This Work&H&23504&8.17 & \\
\hline
WD\,0446$-$255&4881758307940843008&WD\,J044901.40$-$252636.27&\citealt{Swan2019}&He&10120&8.00 & -4.0$\pm$0.1\\
WD\,1350$-$162&6295510766956001024&WD\,J135334.96$-$162706.23&\citealt{Swan2019}&He&11640&8.02 & -5.3$\pm$0.1\\
WD\,1232+563&1571584539980588544&WD\,J123432.68+560643.03&\citealt{Xu2019}&He&11787&8.30 & -5.90$\pm$0.15\\
SDSS\,J1242+5226&1569094249222340352&WD\,J124231.09+522626.61&\citealt{Raddi2015}&He&13000&8.00 & -3.7$\pm$0.1\\
1013+0259&3835858141283862528&WD\,J101347.13+025913.82&\citealt{Izquierdo2023}&He&13158&8.08 & -3.13$\pm$0.01\\
SDSS\,J2339$-$0424&2446993162322393088&WD\,J233917.03$-$042424.67&\citealt{Klein2021}&He&13735&7.93 & -3.51$\pm$0.18\\
SDSS\,J0738+1835&671450448046315520&WD\,J073842.57+183509.71&\citealt{Dufour2012}&He&13950&8.40 & -5.73$\pm$0.17\\
HS\,2253+8023&2286107295188538240&WD\,J225435.71+803953.79&\citealt{Klein2011}&He&14000&8.10 & -5.70$\pm$0.06\\
WD\,1425+540&1608497864040134016&WD\,J142736.17+534828.00&\citealt{Xu2017}&He&14490&7.95 & -4.2$\pm$0.1\\
0944$-$0039&3827999107046766720&WD\,J094431.28$-$003933.75&\citealt{Izquierdo2023}&He&14607&8.76 & -5.87$\pm$0.05\\
Gaia\,J0218+3625&328152307624540032&WD\,J021816.64+362507.60&\citealt{Doyle2023}&He&14691&7.86 & -6.02$\pm$0.15\\
EC\,22211$-$2525&6625468025993283200&WD\,J222358.39$-$251043.57&\citealt{Doyle2023}&He&14743&7.90 & -5.56$\pm$0.15\\
WD\,2207+121&2727904257071365760&WD\,J220934.85+122336.56&\citealt{Xu2019}&He&14752&7.97 & -6.32$\pm$0.15\\
WD\,1551+175&1196531988354226560&WD\,J155409.02+172124.19&\citealt{Xu2019}&He&14756&8.02 & -4.45$\pm$0.08\\
WD\,1244+498&1567391625404274304&WD\,J124703.28+493423.52&\citealt{Doyle2023}&He&15150&7.97 & -5.12$\pm$0.15\\
WD\,1248+1004&3735027667976597248&WD\,J124810.23+100541.22&\citealt{Doyle2023}&He&15178&8.11 & -5.18$\pm$0.15\\
GD\,40&5187830356195791488&WD\,J030253.10$-$010833.80&\citealt{Jura2012}&He&15300&8.00 & -5.10$\pm$0.60\\
G241$-$6&2225982838287401984&WD\,J222333.11+683724.28&\citealt{Jura2012}&He&15300&8.00 & -5.90$\pm$0.30\\
1516$-$0040&4418628372344562048&WD\,J151642.97$-$004042.50&\citealt{Izquierdo2023}&He&15448&8.42 & -4.50$\pm$0.01\\
Gaia\,J1922+4709&2127665711125011456&WD\,J192223.41+470945.37&\citealt{Doyle2023}&He&15497&7.95 & -5.50$\pm$0.15\\
WD\,1145+017&3796414192429498880&WD\,J114833.63+012859.42&\citealt{LeBourdais2024}&He&15500&8.19 & -4.83$\pm$0.14\\
GD\,378&2109852523240471936&WD\,J182337.01+410402.11&\citealt{Klein2021}&He&15620&7.93 & -4.48$\pm$0.15\\
0859+1123&603986308646058496&WD\,J085934.18+112309.46&\citealt{Izquierdo2023}&He&15717&8.19 & -4.84$\pm$0.04\\
0030+1526&2780715106223331072&WD\,J003003.23+152629.34&\citealt{Izquierdo2023}&He&15795&8.18 & -5.01$\pm$0.02\\
0930+0618&3852675309069466496&WD\,J093031.00+061852.93&\citealt{Izquierdo2023}&He&15982&8.18 & -4.87$\pm$0.04\\
1627+1723&4466719602193586816&WD\,J162703.34+172327.59&\citealt{Izquierdo2023}&He&16134&8.29 & -5.05$\pm$0.07\\
1109+1318&3965233688795064832&WD\,J110957.82+131828.07&\citealt{Izquierdo2023}&He&16308&8.25 & -4.01$\pm$0.03\\
SDSS\,J1734+6052&1435966347699773952&WD\,J173435.75+605203.22&\citealt{Doyle2023}&He&16340&8.04 & -4.71$\pm$0.15\\
0259$-$0721&5179698505635671680&WD\,J025934.98$-$072134.29&\citealt{Izquierdo2023}&He&16390&8.26 & -6.04$\pm$0.08\\
GD\,424&547179765520150912&WD\,J022408.71+750257.56&\citealt{Izquierdo2021}&He&16560&8.25 & -3.65$\pm$0.03\\
1359$-$0217&3657725094236611072&WD\,J135933.24$-$021715.16&\citealt{Izquierdo2023}&He&16773&8.14 & -3.16$\pm$0.02\\
J0644$-$0352&3105360521513256832&WD\,J064405.23$-$035206.42&\citealt{Rogers2024}&He&17000&7.98 & -5.2$\pm$0.1\\
GD\,61&203931163247581184&WD\,J043839.37+410932.35&\citealt{Farihi2013}&He&17280&8.20 & -3.89$\pm$0.15\\
WD\,1415+234&1253863445200890112&WD\,J141755.37+231136.71&\citealt{Doyle2023}&He&17312&8.17 & -4.92$\pm$0.15\\
SDSS\,J2248+2632&1883678961315891584&WD\,J224840.93+263251.63&\citealt{Doyle2023}&He&17369&8.02 & -5.09$\pm$0.15\\
WD\,J2047$-$1259&6888044253249541120&WD\,J204713.76$-$125908.94&\citealt{Hoskin2020}&He&17970&8.04 & -1.08$\pm$0.08\\
Ton\,345&689352219629097856&WD\,J084539.18+225728.25&\citealt{Wilson2015}&He&19780&8.18 & -5.1$\pm$0.5\\
WD\,1536+520&1595298501827000960&WD\,J153725.73+515126.83&\citealt{Farihi2016}&He&20800&7.96 & -1.70$\pm$0.15\\
WD\,1622+587&1623866184737702912&WD\,J162259.65+584030.90&\citealt{Rogers2024}&He&21530&7.98 & -3.4$\pm$0.1\\
\enddata
\end{deluxetable*}

\subsection{Calculation of Water Mass Fractions}
We calculate water mass fractions as $f_{\mathrm{H_2O}} = M_\mathrm{{H_2O}/M_{tot}}$ based on the assumption that any excess O in the WD pollution is due to water. In calculating water mass fractions we consider the elements of Ca, Al, Fe, Si, and Mg, and follow \citet{Klein2010} to derive an oxygen budget assuming all elements are present as oxides in the rock. In four cases, one of either Fe or Si is missing. We impose a ratio relative to Mg matching CI chondrite (for example, see Section \ref{section:j0649}). 

Si, Fe, and Mg abundances are typically highest and therefore dominate the budget.
To calculate a \textit{minimum} water mass fraction, we assume all iron is in oxidized form (FeO) as opposed to metal form. Including a metal core would increase $f_{\mathrm{H_2O}}$ as more oxygen would be free to make water. 

We first check for an O excess in each WD by calculating the moles of O required to transform all other abundances into charge-balanced oxide components. If there is an excess of O, the resulting water mass is the molar mass of $\mathrm{H_2 O}$ multiplied by the moles of excess O. Finally, we report water mass fractions by dividing the water mass by the sum of the water mass and mass of the oxides.

\subsection{Atmospheric Accretion and Settling Model}\label{section:juramodel}
To account for effects of the accretion process on elemental abundance ratios and water abundances, we adopt the exponential accretion and settling model by \cite{Jura2009}. This model assumes the disk of parent body material is
completely homogenized (each parcel of disk material reflects the bulk composition of the parent body) and dissipates exponentially as it is
accreted by the WD.  The mass of the disk at a given time $t$ is $M_\mathrm{disk}(t) = M_{\mathrm{PB}} e^{-t/\mathrm{\tau_d}}$, where $\mathrm{\tau_d}$ is the e-folding lifetime of the disk and $M_{\mathrm{PB}}$ is the initial mass of the parent body source of the pollution. 
Typical disk lifetimes are estimated to be $\mathrm{\log \tau_d = 5.6 \pm 1.1}$ \citep{Cunningham2021}.

Pollution settles out of the atmosphere of the WD at a rate of $\dot M_{\mathrm{Z}} = M_{\mathrm{Z}}(t) /\mathrm{\tau_Z}$, where $M_{\mathrm{Z}}$ is the mass of element Z in the atmosphere at a given time and $\mathrm{\tau_Z}$ is a model-derived settling timescale which can vary between days to millions of years for different elements and WD structures \citep{Koester2009, Blouin2018}. 
In the steady state accretion phase the mass flux via accretion is balanced
by elements diffusing out of the WD atmosphere; as noted above
and in \citet{Gansicke2012} this is the most likely state for warm H-WDs.

The fraction of an original body's mass that is in the WD atmosphere at time $t$ is therefore 

\begin{equation}\label{eq:J09_MCV}
     \frac{M_{\mathrm{atm}}(\mathrm{Z},t)}{M_{\mathrm{PB}}(\mathrm{Z})} = \frac{\mathrm{\tau_Z}}{\mathrm{\tau_d} - \mathrm{\tau_Z}} \left(e^{-t/\mathrm{\tau_d}} - e^{-t/\mathrm{\tau_Z}} \right),
\end{equation}
\noindent for each element Z \citep{Jura2009}. According to Equation \ref{eq:J09_MCV}, pollution in the WD atmosphere increases after the onset of accretion, reaches a peak at a steady state point where the accretion and settling rates equilibrate, and then decreases as the accretion disk is depleted and settling dominates the system. We refer to these three phases as the mass-buildup, steady state, and mass-settling phases. 

Accretion reaches the steady state when $dM_{\mathrm{atm}}/dt =0$. Solving for the steady state time, $t_{\mathrm{SS}}$, gives
\begin{equation}\label{eq:J09_tSS}
     t_{\mathrm{SS}}(\mathrm{Z}) = \frac{\mathrm{\tau_d} \mathrm{\tau_Z}}{\mathrm{\tau_d} - \mathrm{\tau_Z}} \ln\left(\frac{\mathrm{\tau_d}}{\mathrm{\tau_Z}}\right). 
\end{equation}

\noindent The steady state point does not occur at exactly the same time for all elements, as it is dependent on the relevant settling timescale $\mathrm{\tau_Z}$. However, the different steady state times are closely spaced, so throughout this work we choose a steady state time that is the median of the $t_{\mathrm{SS}}(\mathrm{Z})$ values for the observed elements for any given WD. 

From Equation \ref{eq:J09_MCV}, we find the \textit{inferred} element ratio of an accreting parent body from an observed abundance ratio to be

\begin{equation}\label{eq:J09_nznnorm}
     \frac{n_1}{n_2} \Big|_{\mathrm{PB}} =  \frac{n_1}{n_2} \Big|_{\mathrm{obs}} \left( \frac{\tau_2}{\tau_1} \right) \left(\frac{\mathrm{\tau_d} - \tau_1}{\mathrm{\tau_d} - \tau_2} \right) \left(\frac{e^{-t/\mathrm{\tau_d}} - e^{-t/\tau_2}}{e^{-t/\mathrm{\tau_d}} - e^{-t/\tau_1}}\right),
\end{equation}

\noindent where $\tau_1$ and $\tau_2$ are the settling timescales for the elements of interest. We use Equation \ref{eq:J09_nznnorm} throughout this work to calculate the compositions of polluting parent bodies from observed abundances. 

For $t \ll \tau_1$, $\tau_2$, and $\mathrm{\tau_d}$, $ \frac{n_1}{n_2} \Big|_{\mathrm{PB}} \approx  \frac{n_1}{n_2} \Big|_{\mathrm{obs}}$; in other words, the observed abundance ratios match those of the parent body at early times
(the mass-buildup phase). At late times in the accretion process
(the mass-settling phase), when $t \gg \tau_1$, $\tau_2$, and $\mathrm{\tau_d}$, the abundance ratios trend towards the ratio of settling times if the disk lifetime exceeds the settling timescales. If the disk lifetime is shorter than the relevant settling timescales, then the parent body abundance ratios diverge towards 0 or infinity. Based on typical settling timescales for H-WDs, and the typical estimated range of $\mathrm{\tau_d}\approx10^5 - 10^7  \mathrm{yr}$, most H-WDs will fall in the limit of $\mathrm{\tau_d} > \tau_1, \tau_2$, while most He-WDs with longer settling timescales will be in the regime of  $\mathrm{\tau_d} < \tau_1, \tau_2$. The steady state phase for H-WDs therefore functionally extends from $t_{\mathrm{SS}}$ onwards. We discuss the implications of these limits for water mass fractions in Section \ref{section:picking_phases}.

\subsection{Bayesian Analysis of Abundances}\label{section:mcmc_methods}

As demonstrated in Section \ref{section:juramodel}, relating observed and parent body compositions requires knowledge of the e-folding lifetime of the accretion disk ($\mathrm{\tau_d}$) and the time $t$ since the onset of accretion at which we observe the WD. We test the effect of selecting different characteristic accretion parameters on elemental abundances and water mass fractions in two ways. First, we manually calculate these quantities assuming mass-buildup, mass-settling, and steady state phases of accretion (results in Section \ref{section:picking_phases}). Separately, we also apply an MCMC approach to explore the most likely accretion parameters for the observed polluted WDs based on the hypothesis that the original bodies were chondritic in composition (Section \ref{section:mcmc_solutions}). This section discusses the details of our MCMC approach. 

\begin{figure}
    \centering
    \includegraphics[width=1.0\linewidth]{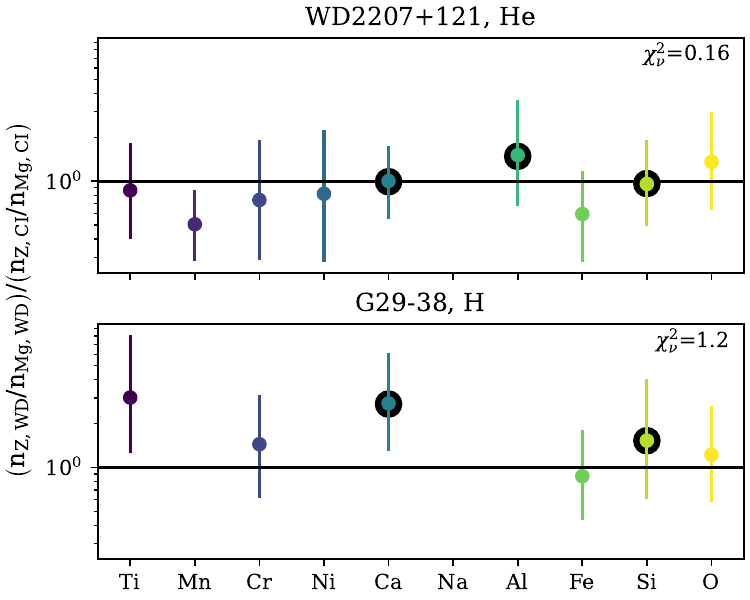}
    \caption{Demonstration 
    of how parent body compositions at a given accretion phase 
    are assessed for agreement with chondritic abundance ratios. 
    Here the mass-buildup accretion phase is shown
    for two WDs in our sample. G29-38 is the coolest H-WD
    in our sample and the only such object 
    that could plausibly be in the mass-buildup
    phase (settling timescales between 1.8-6.5~years; the next coolest H-WD
    is WD\,0145+234 with $<$2 week settling timescales as reported in Table \ref{tab:newHWds_abundance}).
    Ratios are relative to Mg and normalized by CI chondrite abundance ratios. 
    The $\chi^2_\nu$ goodness of fit of the abundance ratios to CI chondrite is listed in the upper right corner, and the error bars show the $2\sigma$ uncertainties used to calculate $\chi^2_\nu$. Only the lithophile elements (outlined in black) are included in the $\chi^2_\nu$ calculation.} 
    \label{fig:WDratios_observed}
\end{figure}

Multiple studies have found that the material polluting WDs is rocky and,
to within typical uncertainties in WD observations,
consistent with chondritic material in the solar system (\citealt{Swan2023, Trierweiler2023, Doyle2023}); Figure \ref{fig:WDratios_observed} provides two examples from the sample in this work. 
Our goal for the Bayesian analysis is thus to identify the combinations of observation times $t$ and disk lifetimes $\mathrm{\tau_d}$ that 
make each WD parent body have the same lithophile abundance ratios as a CI chondrite. Variations in the refractory to volatile (in this case, oxygen) abundance ratios in polluting parent bodies are interpreted as variations in rock/ice ratios and allow us to calculate the most likely water mass fraction.

For each WD we first use simulated annealing, as implemented with the dual annealing package from \textsc{SciPy} \citep{2020SciPy-NMeth}, to perform an initial search as prescribed by Equation \ref{eq:J09_nznnorm} for the $\mathrm{\tau_d}$ and $t$ values that best reproduce the observed abundances of lithophile elements assuming a CI chondrite composition for the parent body. In this search, solutions for most WDs occupy a relatively large area in $\mathrm{\tau_d}$-$t$ space, often spanning several orders of magnitude in one or both dimensions. 

Using the emcee package \citep{emcee}, we then apply an MCMC to the abundances of the lithophile elements for each WD to find the posterior distributions for the accretion parameters. We carry out the MCMC with both a basic log-likelihood function comparing the CI chondrite-derived model abundances to the data (referred to later in the text as the ``Original'' 
distribution) and an amended log-likelihood function which weights the likelihood by the fraction of parent body mass in the atmosphere at the selected time of observation $t$ (Equation \ref{eq:J09_MCV}; referred to later in the text as the ``Weighted'' distribution). 
The second test is predicated on the assumption that we are most likely to observe WD pollution during stages of accretion where the mass of pollution in the atmosphere is at a maximum. Finally, from the MCMC results we derive the distributions of parent body abundances for all elements, recovering the siderophile and atmophile/volatile abundances implied under the assumption of a chondritic rock and the associated water mass fraction.

\section{Results}\label{section:results}

\subsection{Applying diagnostic accretion phases}\label{section:picking_phases}

\begin{figure*}
    \centering
    \includegraphics[width=0.8\linewidth]{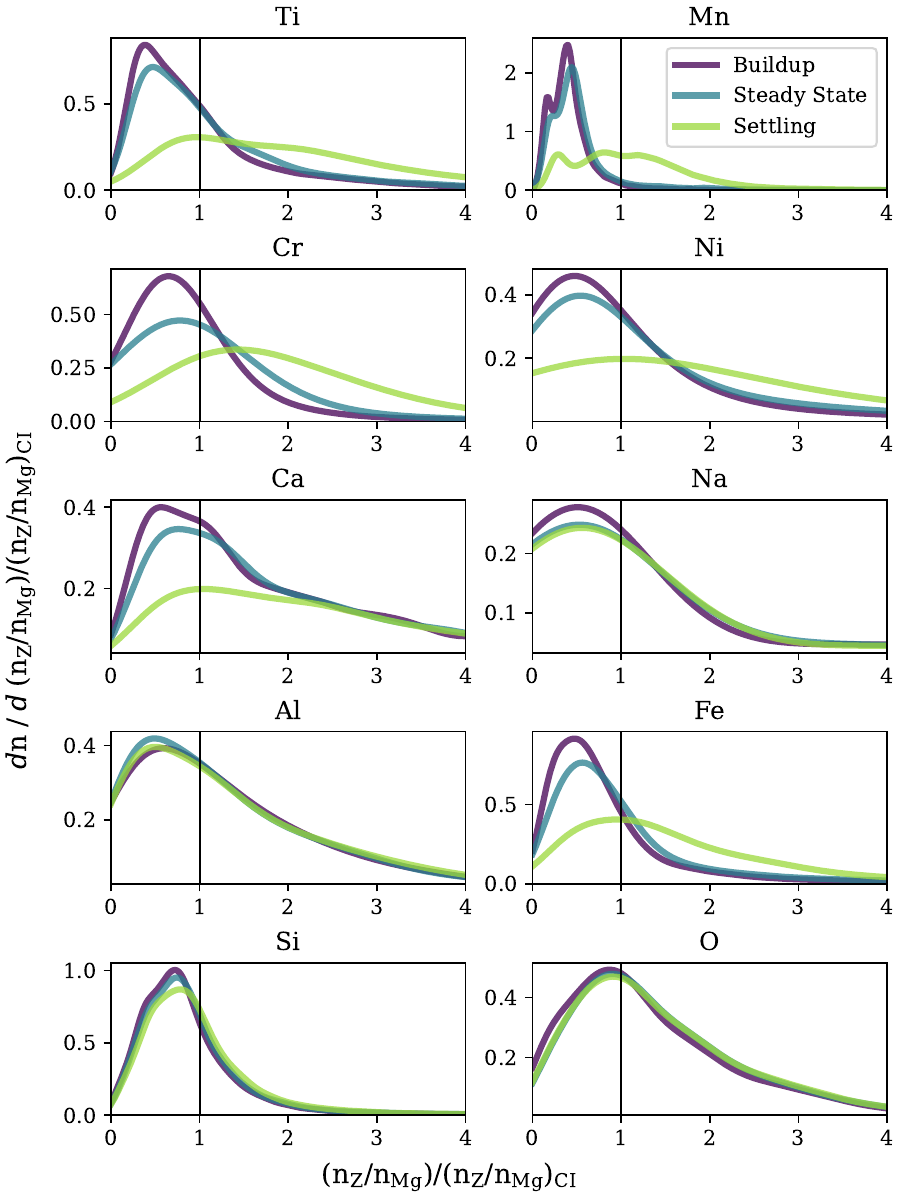}
    \caption{The distributions of abundance ratios calculated for the population of WDs in this study at mass-buildup, steady state, and mass-settling times assuming a disk lifetime of $10^5$ yr. Abundance ratios for Al, Si, and O are largely unchanged at all accretion phases.}
    \label{fig:WD_zdist_phases}
\end{figure*}

Here we carry out the water calculation accounting for various accretion phases. We calculate parent body elemental abundance ratios and water mass fractions across all three phases of accretion (mass-buildup, steady state, and mass-settling) for each WD using Equation \ref{eq:J09_nznnorm}. 

Variations in inferred parent body elemental abundance ratios for a given WD result from adjusting the measured WD atmospheric abundances to different assumed accretion phases and disk e-folding lifetimes.
We consider only the times during which the fraction of parent body currently in the WD atmosphere is non-negligible (Equation \ref{eq:J09_MCV}). 
As discussed in Section \ref{section:juramodel}, inferred parent body ratios in the mass-settling phase will reach different limits depending on the ratio of settling timescales and disk lifetimes. 
In the He-WD case, the relatively long settling timescale for oxygen leads to O abundances diverging later than most other elements. 

To illustrate the net effect of the evolution of parent body abundance ratios with time on a population level, we calculate the  distribution of elemental abundances summed across all WDs in our sample at each phase of accretion. For this purpose, we select a diagnostic time within each accretion phase at which to calculate values, though it should be remembered that abundances may change fairly drastically over the full course of each phase. The two diagnostic times for the mass-buildup and mass-settling phases are $t_{\mathrm{buildup}} = 0.5 \times \min ( \tau_{d}, \tau_{z})$ and $t_{\mathrm{settling}} = 2 \times \max ( \tau_{d}, \tau_{z})$, respectively. The steady state times, $t_{\mathrm{SS}}$, are given by Equation \ref{eq:J09_tSS} and are element dependent, so we calculate a steady state time for each observed element and take the median as our assumed value. 

Figure \ref{fig:WD_zdist_phases} shows the parent body elemental abundance ratio distributions (summed over all WDs) at the mass-buildup, steady state, and 
mass-settling diagnostic times with a disk lifetime of $10^5$ yr. While the parent body ratios calculated in the mass-buildup and steady state phases generally match each other, the settling phase solutions typically diverge as described above.
The distribution of abundance ratios is drawn somewhat closer to chondritic values in the mass-settling phase; however, the spread in the distribution increases dramatically, reflecting the sensitivity of the mass-settling phase values to the exact observation time considered for each WD.

\begin{figure}
    \centering
    \includegraphics[width=1\linewidth]{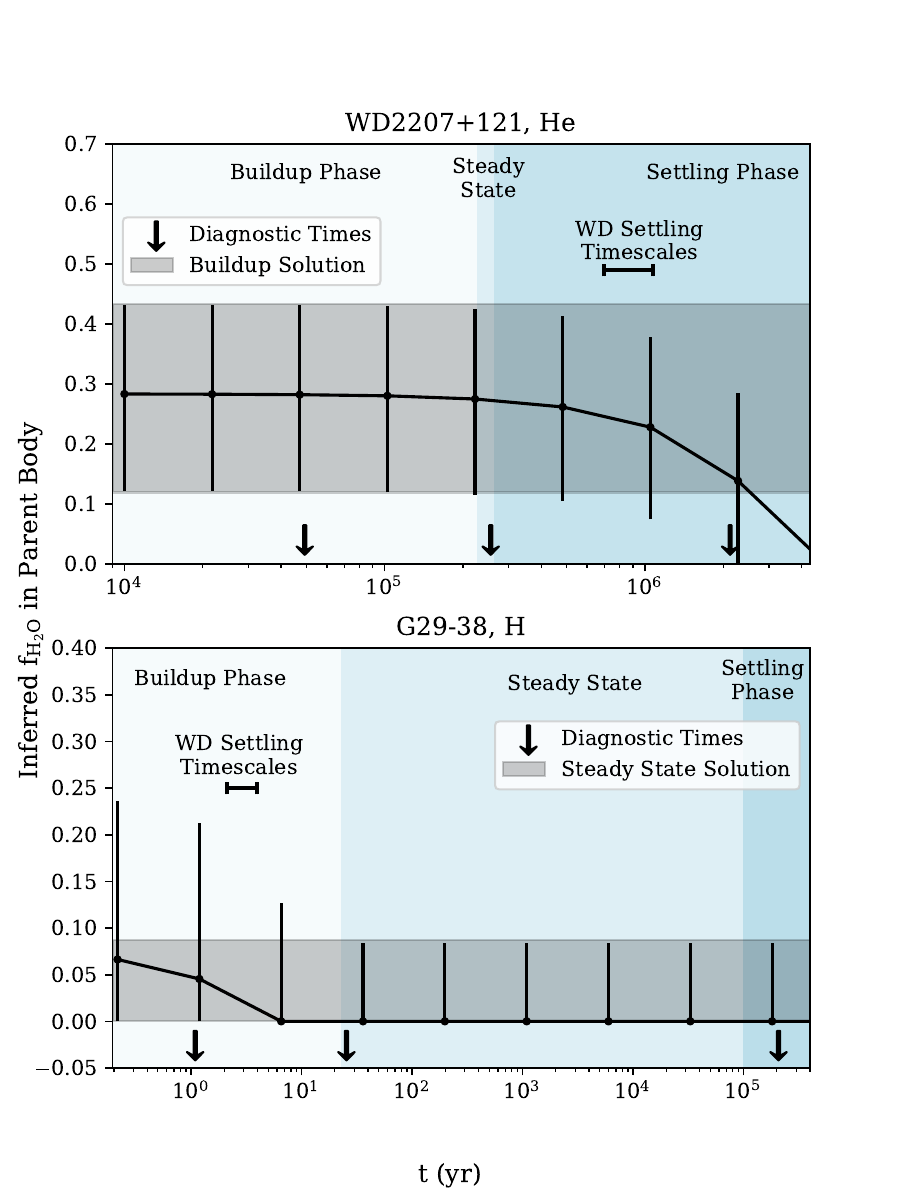}
    \caption{Evolution of the inferred water mass fraction calculated over time for two example WDs assuming a disk lifetime of $10^5$ yr. Water fractions with propagated errors are shown at various times during the evolving accretion process. The shaded grey region shows the $ 1\sigma$ water mass fraction calculated for the accretion phase indicated in each plot panel. The arrows show the times at which we sample water mass fractions in the mass-buildup, steady state, and mass-settling phases of accretion. The ranges of settling timescales for each WD are shown by the horizontal bars.}
    \label{fig:water_phases_time}
\end{figure}

Figure \ref{fig:water_phases_time} shows an example of calculating
$\mathrm{f_{H_2O}}$ from the inferred parent body elemental abundances
at each phase of accretion for two example WDs previously shown in 
Figure \ref{fig:WDratios_observed}; a disk lifetime of $10^5$ yr
is assumed. The error bars show the uncertainties in water mass fraction obtained by propagating the uncertainties in the measured atmospheric elemental 
abundances through Equation \ref{eq:J09_nznnorm}. As a result of the mathematical limits of the accretion process (Section \ref{section:juramodel}), water mass fractions either stabilize or diverge towards zero in the settling phase.

\begin{figure}
    \centering
    \includegraphics[width=1\linewidth]{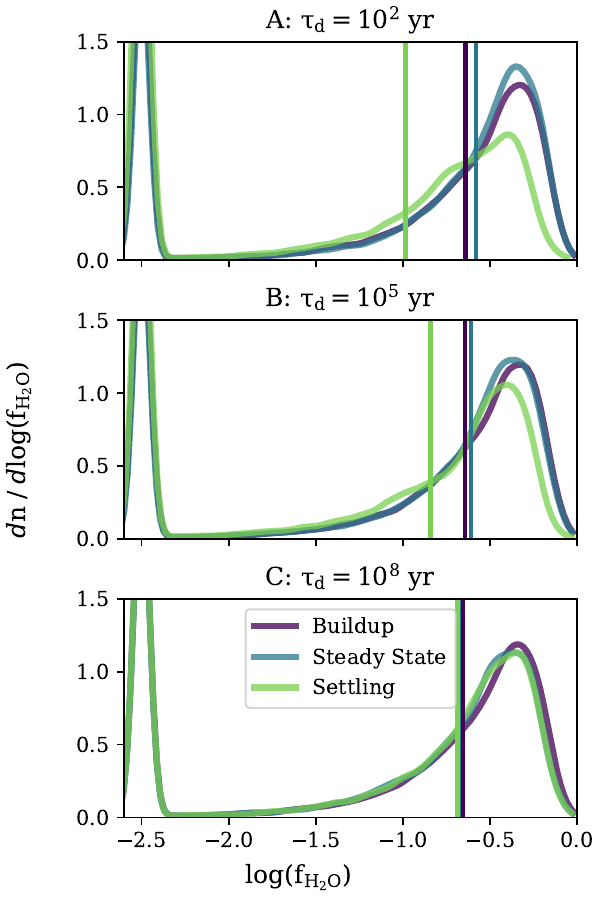}
    \caption{Water mass fractions calculated for the mass-buildup, steady state, and mass-settling phases assuming three different disk lifetimes. The median value for each curve is indicated by the vertical lines. For visualization, water fractions of zero are reported as $10^{-2.5}$. Unless the disk lifetime is very short (panel A) compared to the settling timescales, the water mass fractions in the mass-buildup and steady state phases are very similar. The inferred parent bodies are more likely to present as dry in the mass-settling phase.}
    \label{fig:water_phases}
\end{figure}

We now calculate the summed distribution of $\mathrm{f_{H_2O}}$ for the full sample of WDs derived at the mass-buildup, steady state, and mass-settling diagnostic times assuming $\mathrm{\tau_d}=10^5$ yr (panel B of Figure \ref{fig:water_phases}).
We emphasize the bimodal distribution defined by ``dry" vs ``non-dry" accreted bodies  by artificially placing pollution with $0\%$ water at $\mathrm{f_{H_2O}} = 10^{-2.5}$. Inferred water mass fractions typically decrease in the settling phase as the majority of WDs in our sample are He-WDs. For He-WDs, the low inferred water fractions in the settling phase are largely due to parent body Fe/Mg increasing exponentially at late times, decreasing O excesses. 

Thus far we have considered a single $\mathrm{\tau_d}$; we now vary $\mathrm{\tau_d}$ to illustrate how disk lifetimes much longer or shorter than $\sim10^{5}$ yr would alter inferred water mass fractions, based on Equation \ref{eq:J09_nznnorm}. 
Panels A and C of Figure \ref{fig:water_phases} show the summed distributions of water mass fractions across the whole WD sample at each phase of accretion for short and long disk lifetimes of $10^2$ and $10^8$ yr, respectively. We find minimal changes in water distributions between different disk lifetimes in the buildup and steady state phases. A short disk lifetime leads to inferring more of the pollution to be dry in the mass-settling phase. Conversely, a long disk lifetime that exceeds settling timescales and allows elemental abundance ratios to reach stable limits will keep water mass fractions close to the mass-buildup and steady state values (panel C).

\subsection{Accretion phases derived from MCMC}\label{section:mcmc_solutions}

As described in Section \ref{section:mcmc_methods}, we employ an MCMC procedure to 
constrain the most likely phase of accretion, defined by $t$ (the time since the onset of accretion at which the WD is observed) and $\mathrm{\tau_d}$ (the e-folding lifetime of the accretion disk), for each WD in our sample assuming
each parent body has ratios of lithophile elements matching those of CI chondrites. We start with generous bounds of $10^{-6}$ to $10^{10}$ yr for both accretion parameters in our search. We then run these tests again with a prior restricting the disk lifetime to $10^4 - 10^7$ yr (current disk lifetime estimates are $\mathrm{\log \tau_d = 5.6 \pm 1.1}$, \citealt{Cunningham2021}) and refer to these results as ``Restricted $\mathrm{\tau_d}$".

We apply two different likelihood functions: the first (referred to as ``Original'') calculates a standard log likelihood comparing the observed abundances to the modeled abundances based on the accretion parameters and assuming a chondritic parent body. The second (referred to as ``Weighted") weights the fit by the fraction of the parent body mass that would be in the atmosphere of the WD at the time $t$. This aims to disfavor solutions which imply very small fractions of the parent body being present, resulting in anomalously high parent body masses.

\begin{figure}
    \centering
    \includegraphics[width=1\linewidth]{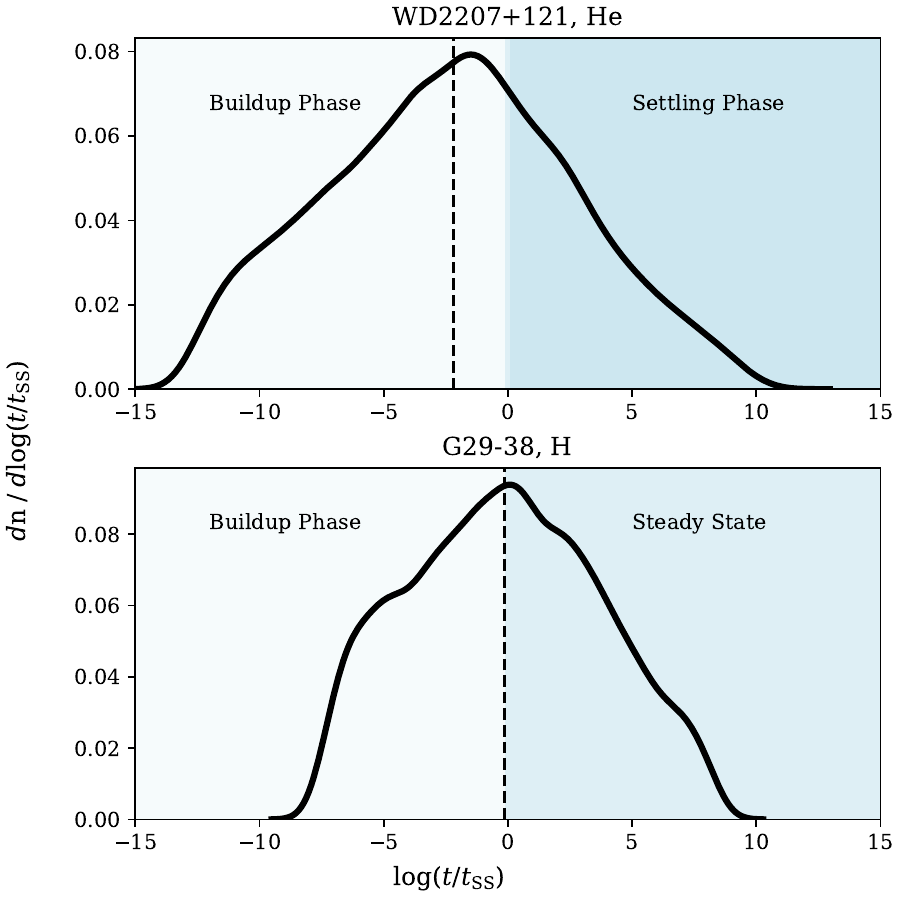}
    \caption{The distribution of inferred observation times relative to steady state 
    ($t_{\mathrm{SS}}$, calculated from $\mathrm{\tau_d}$ and the median settling time through Equation \ref{eq:J09_tSS}) for two example WDs, derived from the posteriors of the MCMC test which uses a standard likelihood function and no strong prior on the disk lifetime
    (the ``Original'' likelihood function). Solutions to the left of the vertical line correspond with the mass-buildup phase (light shaded region) while the right (darker shaded region) is generally the mass-settling phase for the He-WDs and typically samples the steady-state for H-WDs. The median of each WD's distribution is shown as the vertical dashed line. Most WDs have a broad distribution with a peak around steady state. }
    \label{fig:MCMC_t_tSS}
\end{figure}

Without placing strong priors on the disk and observation times, we find that the distribution of accretion parameters obtained from the MCMC is usually very broad. This is because abundances typically only vary significantly at late or early times during the accretion process, depending on the relative disk lifetimes and settling timescales (see Section \ref{section:picking_phases}), leading to long periods of time where abundance ratios are relatively constant. 
Figure \ref{fig:MCMC_t_tSS} shows the ``Original" likelihood function
posterior distributions from the MCMC test, recast as $t/t_{\mathrm{SS}}$, for the two example WDs from Figure \ref{fig:WDratios_observed}. 

\begin{figure*}
    \centering
    \includegraphics[width=1\linewidth]{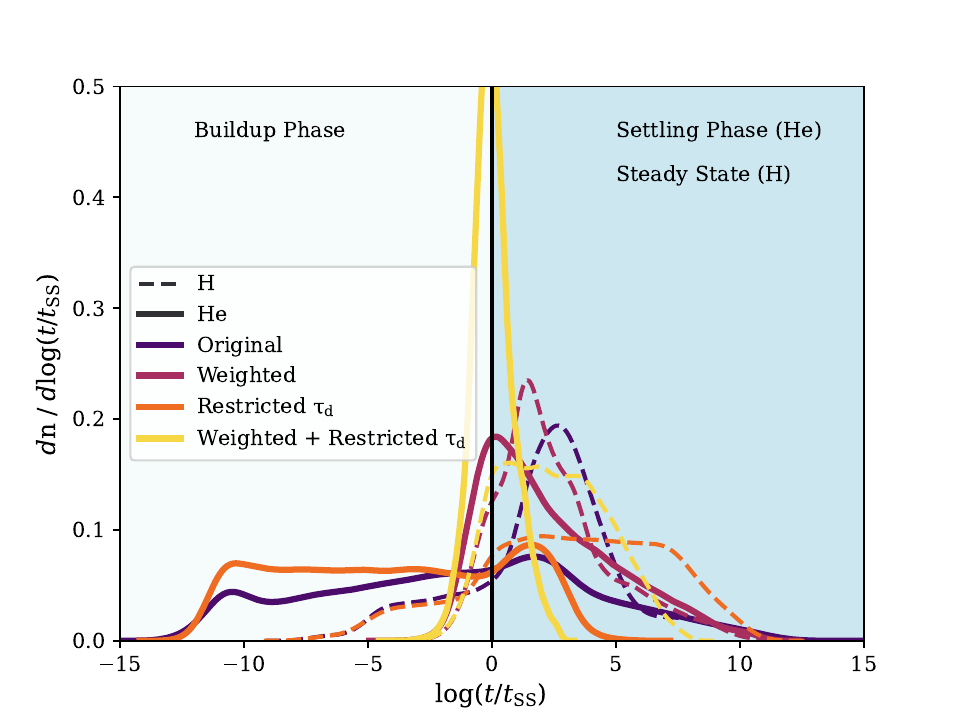}
    \caption{The distribution of accretion parameters relative to steady state derived from the posteriors of the MCMC tests, separated based on the dominant atmospheric species. To compare accretion phases, we have recast $\mathrm{\tau_d}$ as $t_{\mathrm{SS}}$ according to Equation \ref{eq:J09_tSS}.
    Solutions to the left of the vertical line correspond with the mass-buildup phase while the right is generally the mass-settling phase for the He-WDs and typically samples the steady-state for H-WDs. Most of the individual white dwarfs peak around steady state, and have broad distributions. ``Weighted" refers to applying a likelihood function that prefers solutions that maximize the mass of pollution in the atmosphere. ``Restricted $\mathrm{\tau_d}$" curves place a restricted prior on the disk lifetime of $10^4-10^7$ yr, while non-restricted curves allow $\mathrm{\tau_d}$ to vary freely between $10^{-6}$ to $10^{10}$ yr. Applying a weighted likelihood function to prefer solutions that maximize the mass currently in the atmosphere of the white dwarf tends to bring solutions towards the steady state and mass-buildup phases. H-WDs tend to peak at relatively later accretion parameters than He-WDs but for relevant settling timescales still easily reside in the
    steady state.}
    \label{fig:MCMC_t_tSS_all}
\end{figure*}

To show the effects of the different MCMC procedures, we sum all of the $t/t_{\mathrm{SS}}$ distributions across the WD sample (segregating H- and He-WDs) for each MCMC test and show the summed distributions in Figure \ref{fig:MCMC_t_tSS_all}. Tests include running the MCMC with the standard
(``Original'') and weighted (``Weighted'') likelihood functions and with (``Restricted $\mathrm{\tau_d}$'') and without a restricted prior on $\mathrm{\tau_d}$. H-WDs tend to have posterior distributions that peak at steady state, consistent with the hypothesis that we should observe H-WDs while they are actively accreting material due to their extremely short settling timescales. He-WDs have a much flatter distribution across accretion parameters, though applying the weighted likelihood function drives the MCMC solutions towards steady state, as that is when the mass in the atmosphere is at a maximum. This reflects how abundances for the He-WDs are often unchanged from the buildup to steady state phases, and begin to diverge in the settling phase (Section \ref{section:picking_phases}). 

Amongst the individual objects in our sample, we find that WDs with peaks in the distributions of $t/t_{\mathrm{SS}}$ in the settling or near-steady state phases remain the same when the weighted likelihood is applied, but three WDs with mass-buildup phase solutions move to steady state. Placing a prior on $\mathrm{\tau_d}$ of $10^4 - 10^7$ yr tends to flatten the distributions in Figure \ref{fig:MCMC_t_tSS_all}, with most of the values on the buildup side. Applying the weighted likelihood function in addition to the more restricted prior on $\mathrm{\tau_d}$ brings the posteriors back to be centered on steady state, again with the exception of any WDs whose distributions peak in the mass-settling phase.  

These tests are based on the assumption that the true parent body abundances of each pollution sample are chondritic. To assess that assumption, we use the chain of $t$ and $\mathrm{\tau_d}$ from the MCMC and the observed lithophile abundances for the WDs to recover a set of parent body abundances at each draw. We then calculate the $\chi^2_\nu$ for the fit of these abundances to CI chondrite and find that 36/51 of the WDs have $\chi^2_\nu$ distributions for which the majority of the distribution is lower than the critical value for the body passing as chondritic. Applying the weighted likelihood function and/or placing a more restricted prior on $\mathrm{\tau_d}$ does affect the $\chi^2_\nu$ distributions, but only moves a handful of WDs from passing to not passing as chondritic, or vice versa. 

\begin{figure}
    \centering
    \includegraphics[width=1\linewidth]{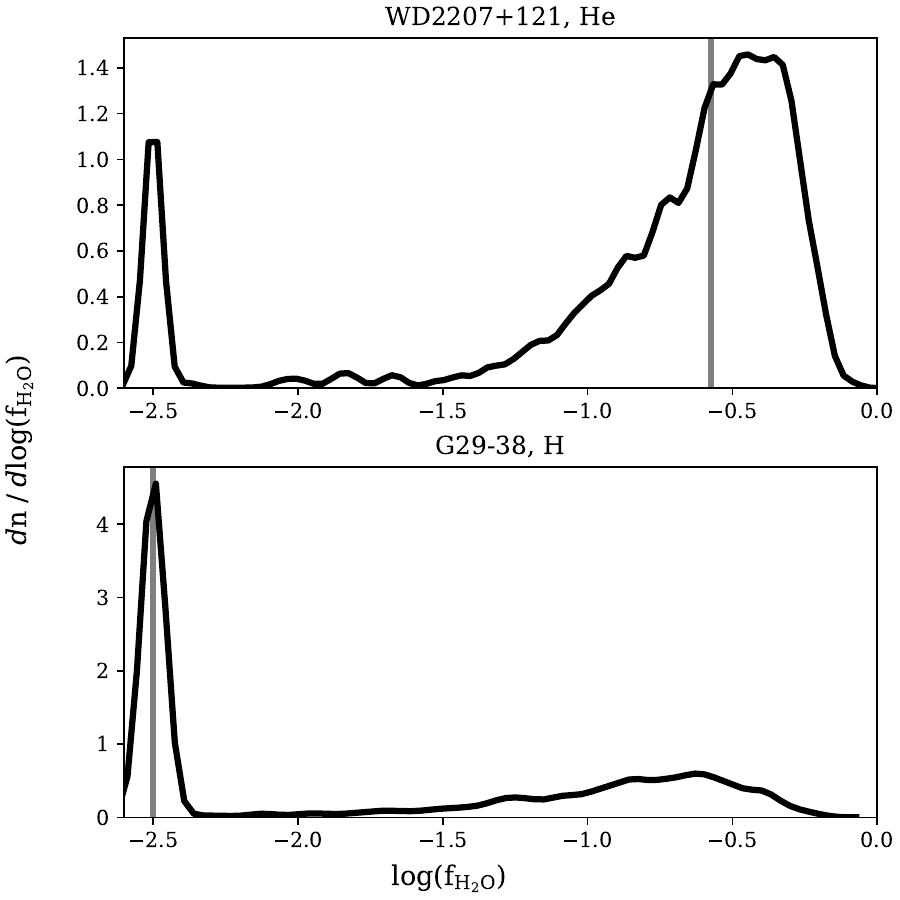}
    \caption{The distributions of water mass fractions derived for two example WDs, based on the observed abundances and the accretion parameter posteriors of the MCMC test for the ``Original'' likelihood function. For visualization, water abundances of zero are plotted at $10^{-2.5}$. The median $\mathrm{f_{H_2O}}$ is shown as the vertical line.}
    \label{fig:MCMC_water}
\end{figure}

Finally, we use the inferred parent body elemental abundances derived from the MCMC chain to calculate distributions of $\mathrm{f_{H_2O}}$ for each WD. Figure \ref{fig:MCMC_water} shows the individual water distributions for our two
example WDs from Figure \ref{fig:WDratios_observed} for the ``Original'' MCMC test. As previously, we set all water mass fractions of 0 to $10^{-2.5}$. 
The posterior distributions of parent body $\mathrm{f_{H_2O}}$ are very broad for most WDs, except for those with no evidence for water (``dry"). As the MCMC chain often spans different phases of accretion, the resulting distribution of water fractions will often include both ``dry" and ``wet" solutions. As shown in Section \ref{section:picking_phases}, water mass fractions quickly move towards zero after a few settling timescales. This leads to a buildup of dry solutions for any posterior distributions that sample the settling phase and results in a bimodal shape in the log of the water fraction.

When reporting values for a single WD, we take the median and $1\sigma$ values of the $\mathrm{f_{H_2O}}$ distribution. Applying a weighted likelihood function minimally changes the shape of the water mass fraction distribution for each WD, and the peaks of the distributions do not change when steady state is preferred by the weighted likelihood. Placing a restricted prior on $\mathrm{\tau_d}$ similarly has a minimal impact on the water mass fraction peaks and overall distributions. 

\section{Discussion}\label{section:discussion} 
\subsection{Implied accretion phases for H and He WDs}

In this section we discuss which accretion phases were generally selected
for H- and He-WDs during the MCMC tests and how that compares
to expectations for each class as discussed earlier in the text.

H-WDs, especially those considered in this paper, are unlikely to be caught
in anything other than the steady state accretion phase.
The results from the MCMC tests support this statement (Figure \ref{fig:MCMC_t_tSS_all}). For individual WDs, using the parent body elemental abundances solved over time for a disk lifetime of $10^5$ yr, we find that the fit to chondrite generally improves around the steady state point for H-WDs, though the improvement is usually not enough to bring the abundance ratios into agreement with chondrites. 

For WDs with longer settling times, such as most He-WDs, systems typically need to reach a mass-settling phase before significant changes are expected in the calculated parent body abundance ratios.
For the He-WDs, we find that only a few can be forced to fail the $\chi^2_\nu$ test to chondrites when it is assumed they are in the late mass-settling phase (at 5-10 times a typical settling time for that WD). In other words, the assumption of particular accretion parameters does not strongly influence the elemental abundance ratios as long as disk lifetimes are still shorter than the settling timescales.

Only a few He-WDs yield chondritic abundance ratios through the MCMC search for accretion parameters in the mass-settling phase. The summed MCMC results for the WD sample overall (Figure \ref{fig:MCMC_t_tSS_all}) show a small peak in the posterior distributions in the settling phase soon after steady state, but the distribution is very broad and includes a long tail into the buildup phase. Overall, there is a relative robustness of abundances matching chondritic
(relative to typical uncertainties in the observed abundances)
until late in the settling phase. Considering the extreme sensitivity of calculating parent body compositions at late times to the assumed time of observation, we suggest that He-WD abundances are best used in the mass-buildup phase with the knowledge that there may be some outlier WDs at very late accretion phases.

For the remainder of the discussion in this paper we assume
H-WDs are accreting in the steady state and He-WDs
the mass-buildup phase.

\subsection{The distribution of water abundances and water sensitivity}
\label{section:discussionsub1}
We find a wide range of water mass fractions ($\mathrm{f_{H_2O}}$) for the WDs.
Median water abundances for each of the methods described throughout this work along with the 16th and 84th percentiles are reported in Table \ref{tab:water_summary}. 

\begin{figure*}
    \centering    
    \includegraphics[width=1\linewidth]{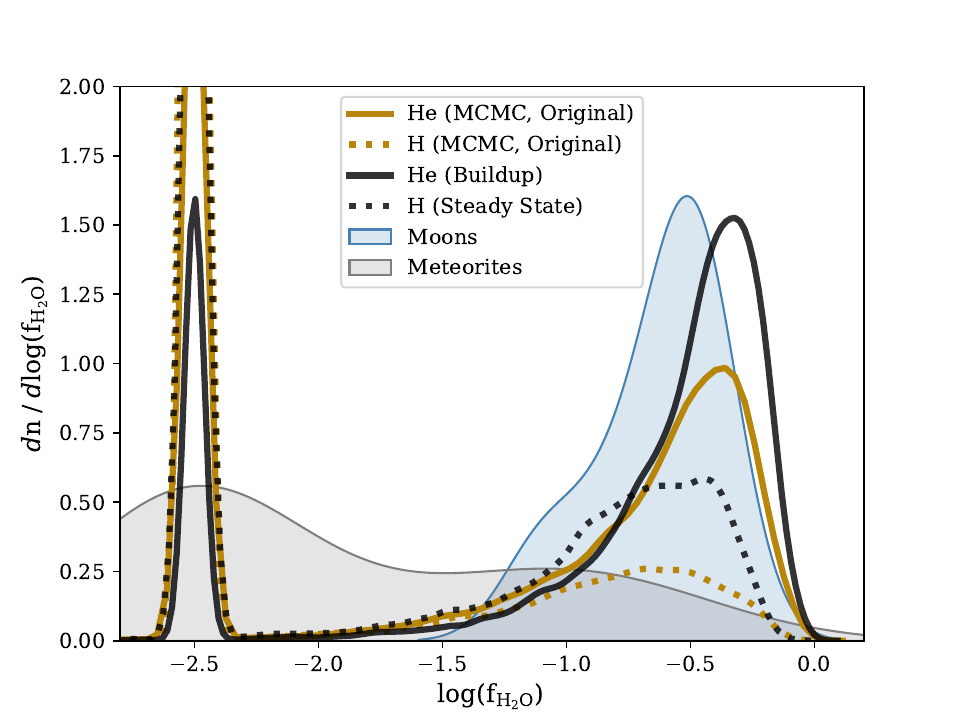}
    \caption{The summed water mass fraction distribution for the WDs from the ``Original" MCMC test for the H and He WD samples, under the assumption of chondritic parent bodies, compared to the buildup (He WDs) and steady state (H WDs) results. For clarity, we do not show the results from additional MCMC tests; these results are listed in Table \ref{tab:water_summary} and are typically indistinguishable from the ``Original" curve. Distributions of water mass fractions of meteorites and solar system moons are shown as the two shaded regions. The distribution of meteorites is largely water-poor while the distribution of solar system moons is more water-rich (see Section \ref{section:discussionsub1}).
    }
    \label{fig:water_solution_total}
\end{figure*}

For both H- and He-WDs, water abundances tend to be similar regardless
of the accretion phase selected or the MCMC test executed. 
In any case, as discussed above,
we expect all H-WDs to be in the steady state.
For He-WDs where different accretion phases could be obtained, Fe abundances appear to have the strongest effect on water mass fraction due to the relative abundance of Fe changing especially drastically at late times. 
However, we do not expect many of the He-WDs to be in the mass-settling
phase and indeed instead expect that they should generally be in the
mass-buildup phase.

Before embarking on an assessment of the water mass fraction
values reported in Table \ref{tab:water_summary} it is important
to determine the sensitivity of our method.
Examining the uncertainties quoted for each type of WD
(H- vs He-dominated) and the most likely accretion phase
(steady state and mass-buildup, respectively),
we see that water mass fraction uncertainties can range from 5-30\%
with a median of 15\%. There is a subpopulation of sources that never have excess O, resulting in $\mathrm{f_{H_2O}} = 0\%$ with no uncertainty; this accounts for about 13\% of the sample of 51 WDs considered.
The take-away from this is that given the current capabilities in modeling
WDs it is generally not possible to assess water mass fractions
at a level better than 15\%, but it is possible to say with confidence
if a parent body is inescapably dry. We do not use this metric to test whether single objects are water-rich, but rather to point out that while we can generally differentiate between completely dry and extremely water-rich bodies, we cannot probe water abundances at lower, Earth-like levels.

Figure \ref{fig:water_solution_total} shows the calculated water mass fraction distributions for the WD population in this study compared to those derived from thermal models for solar system moons by \cite{Reynard2023} and those calculated from meteorite abundances \citep{Lodders2021}. Most of the meteorites are dry, with the exceptions of CI and CM which have inferred $\mathrm{f_{H_2O}}$ of approximately $25\%$ and $15\%$, respectively, due to the presence of phyllosilicate minerals produced by aqueous alteration. CK and CR chondrites also have non-negligible water abundances of a few percent. Considering the sample as a whole, and using a combination of buildup (He) and steady state (H) abundances,
we find that about a third of the WDs appear to have accreted dry parent bodies (median $\mathrm{f_{H_2O}} = 0\%$), with the rest having accreted rocky bodies with median water mass fractions of about $\mathrm{f_{H_2O}} = 30\%$, on par with estimates for asteroid 1 Ceres in the solar system \citep[e.g.,][]{deSanctis2015}. The MCMC results return dry compositions for a larger fraction of bodies, however the median $\mathrm{f_{H_2O}}$ of the non-dry bodies remains similar at $\approx25\%$. Despite the significant uncertainties for the WD parent body 
water mass fractions, the overall distribution is consistent with the picture of WD pollution as a mix of dry, rocky material like asteroids and more icy/watery bodies such as moons or water-rich asteroids.

\subsection{Water inferred from H versus excess O}\label{section:HvsO}

H has often been used as a tracer for water abundances in WD pollution under the assumption that H gradually builds up in a WD atmosphere over the course of water accretion \citep[e.g.,][]{Jura2012b,GentileFusillo2017}. Water abundances are then estimated from the accretion rate of H relative to that of all other metals in the WD atmosphere, assuming that the accretion rate of H is the total mass of H in the atmosphere divided by the cooling age of the WD. Our analysis using O excesses as a tracer for water results in somewhat higher water mass fractions than are obtained from H accretion (reaching up to tens of percent by mass, as opposed to a few percent). This difference arises because obtaining water fractions via H and metal accretion rates compares cumulative H accretion to an instantaneous metal accretion rate. As pointed out by \citet{GentileFusillo2017}, because H does not settle it is impossible to tell how much of the H accreted with the observed metals. Inferred water contents can therefore vary greatly depending on whether we allocate a lifetime-averaged amount of H to the body versus assuming the whole of the H atmosphere was accreted with the ongoing pollution event. 

Oxygen provides an instantaneously-measured counterpart to H-derived water abundances, and we can use H abundances as a sanity-check for the water fractions derived from O excesses. Assuming all H in a He-WD atmosphere is either primordial or accumulated over the course of water accretion, there should be at least enough H to convert all excess O to water. For each of the He-WDs, we compare the calculated distribution of excess O abundances in the parent bodies to the observed H abundances to check for agreement. We calculate both H and excess O relative to Mg, propagating the uncertainties from all elements. We find that 30/39 of the He-WDs have distributions of H abundances where the median moles of H are at least large enough to account for the median moles of excess O as being due to water. Four additional WDs have median H abundances that are within 1$\sigma$ of the median excess O (Gaia\,J0218+3625, G241-6, 0944$-$0039, and HS\,2253+8023). 

Finally, five He-WDs have such low H abundances that the moles of H are never sufficient to transform the moles of excess O to water (SDSS\,J0738+1835, 0259$-$0721, 0930+0618, WD\,2207+121, and WD\,1232+563). This most likely suggests an issue in atmospheric modeling for these stars (see Section \ref{sec:3.1}), though it is also possible that O is accreted as a species other than water, such as some molecule containing carbon. Selecting very short disk lifetimes $\ll10^5\mathrm{yr}$ can reduce the ratio of excess O to $\mathrm{H}/2$ by up to a factor of a few; except for SDSS\,J0738+1835, the ratios can be minimized to less than a factor of ten near steady state and early-decreasing phases. Variations in settling time models could similarly impact abundances. Finally, element abundances are dependent on the mass of the convection zone for each white dwarf; uncertainties in modeling can therefore also affect the ratio of H to excess O \citep[e.g.,][]{Dufour2007}. 

The five WDs lack observations in the UV to place limits on C abundances, and we therefore require further observations to test the likelihood of carbon as a contributor to the oxygen budget. We calculate the moles of C required to make $\mathrm{CO_2}$ from the observed O, then use the modeled He atmosphere mass for each WD to calculate the associated $\mathrm{\log C/He}$ abundance. We find that abundances of $\mathrm{\log C/He}\gtrsim-6$ would be required to convert all of the excess O to $\mathrm{CO_2}$, values that should typically be detectable in the UV for WDs. 

\begin{figure*}
    \centering
    \includegraphics[width=1\linewidth]{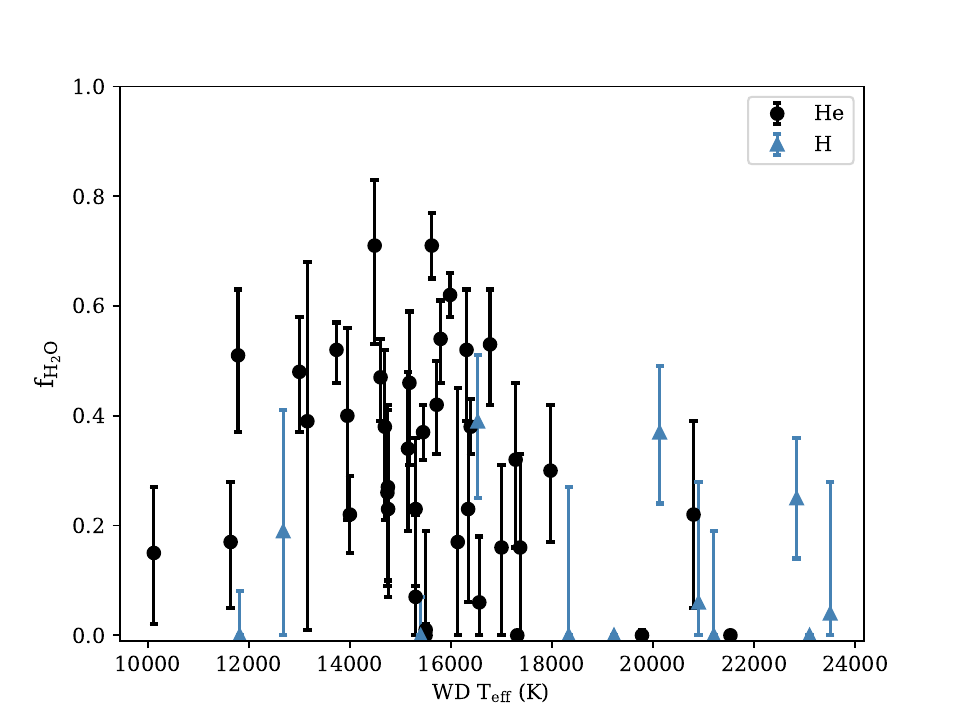}
    \caption{Water mass fractions derived from observed pollution abundances versus the temperature of the WD; H-WDs are corrected assuming steady state
    while He-WDs with the mass-buildup accretion phase. High temperature WDs, including most of the H-WDs, tend to have lower water mass fractions.}
    \label{fig:trend_tests}
\end{figure*}

\subsection{Water History}\label{section:evaporation}
In this work, we consider effects of accretion and settling on derived parent body abundances. However, there are a few processes prior to accretion that could skew inferred water abundances from the real distribution of exoplanetary material. For example, \cite{Malamud2017} suggest that evaporation during the post-main sequence evolution of planetary systems could alter water abundances, leading to trends in water mass fraction with main sequence stellar masses. They argue that more water can be retained by bodies orbiting lower-mass progenitor stars, leading to higher water fractions for low mass WDs. Using the values from Table \ref{table:WDsample}, we do not find a correlation between water mass fractions and WD masses, however updated, consistently-modeled WD masses using {\it Gaia} parallaxes are required to properly test for a correlation.

\cite{Jura2010} suggest water retention rates should also vary with the initial size of the body, with smaller objects drying out more quickly during post-main sequence evolution. While we do not find a correlation between water mass fractions and WD mass, it is feasible that our sample, which represents some of most highly polluted WDs, could be sampling larger bodies that are particularly efficient at retaining water throughout the post-main sequence and accretion processes. This would be consistent with the relatively large parent body masses required to explain most highly polluted WDs \citep{Trierweiler2022}. While total accreted mass does generally increase with inferred $\mathrm{f_{H_2O}}$, the trend is not statistically significant in our sample (p-values $> 0.4$ for both the He and H samples). 

Additionally, models for asynchronous accretion of WD pollution suggest the timing of sublimation and water accretion is determined by the temperature of the WD \citep{Malamud2016, Brouwers2023}. \citet{Brouwers2023} specifically predict higher temperature WDs should have lower water abundances. Figure \ref{fig:trend_tests} shows the observed water mass fractions plotted as a function of WD temperature. Calculating the Pearson correlation coefficient, we find a negative correlation of $-0.31$ for the He-WDs with a p-value of $0.06$. We also find a weak negative correlation of $-0.04$ amongst the H-WDs, however it is not significant (p-value of $0.91$). 
While there does seem to be a lower water incidence rate for hotter WDs (T$_{\rm eff}$$\gtrsim$20,000\,K; 38\% wet in a sample of 8 which includes
only 2 He-WDs),
the sample size remains small and incidence rates and water mass fractions
are not immediately incompatible with cooler 
(10,000\,K$\lesssim$T$_{\rm eff}$$\lesssim$20,000\,K; 77\% wet in a sample of 43) WDs. Given the difference in temperature coverage of the H and He samples, and lack of significant negative trend in the H sample, there does not appear to be clear evidence in support of asynchronous accretion. 
If the effects of post-main sequence evaporation and/or two-phase accretion are shaping the distribution of inferred water abundances, we anticipate that correcting for these effects would shift both the extreme high and low ends of the water distribution towards moderate water fractions.

It is worth mentioning the apparent disagreement between results
presented here for water incidence rates and those found in
\citet{Brouwers2023}. Within the population of He-WDs, \citet{Brouwers2023}
find very few systems that they conclude are accreting water-rich parent 
bodies while our work implies a sample of polluters that skew towards non-dry compositions. One contributing factor
to this
disagreement is how one interprets the results of water mass fraction/oxygen excess distributions. Indeed, both \citet{Brouwers2023} and our own
analysis arrive at similar distributions for water contents
(Figure D2 of \citealt{Brouwers2023} and Figure \ref{fig:MCMC_water}
for our analysis). A significant difference is that \citet{Brouwers2023}
only allow systems with $>$2$\sigma$ oxygen excesses to be labeled as
water-rich whereas our work seeks to describe the trend in median water mass fractions without categorizing each individual body as wet or dry (see discussion on water mass fraction sensitivity above). 

If we adopted similarly restrictive oxygen excess requirements as \citet{Brouwers2023}, 
we would conclude that overall $\approx$60\% of our sample
is dry and $\approx$40\% have accreted detectable amounts of water.
While this assessment 
would lead us to adopt a general conclusion similar to \citet{Brouwers2023},
the revised water incidence rate would still be substantially higher than
that suggested by \citet{Brouwers2023} of $\approx$94\% dry and $\approx$6\% wet. Even increasing our threshold to a $3\sigma$ requirement on water mass fractions, we find that $\approx30\%$ (1 H-WD and 15 He-WDs) are wet.

To test the hypothesis that the majority of the pollution should be dry instead of wet, we ran a second set of MCMC tests in which the likelihood function minimized the amount of excess O for each WD. We found that for both H- and He-WDs, reducing a measured O excess required assuming a late phase of accretion, resulting in huge amounts of Ca and Fe in the parent body (at least 10 times chondritic values). This implies that if all polluting material is dry, then we require extremely non-chondritic parent body abundances, contradicting the typically bulk Earth/chondritic compositions observed.

\subsection{H- vs He-WDs}\label{section:HvsHe}

We find an important distinction between the water characteristics of pollution on H- and He-WDs. Overall, 6/12 of the H-WDs have median $\mathrm{f_{H_2O}}$ values of zero at steady state while 4/39 He-WDs have median $\mathrm{f_{H_2O}}$ values of zero in the buildup phase. 
The number of He-WDs with well-characterized atmospheric pollution
and oxygen detections greatly outnumbers the H-WDs.
H-WDs in general tend to be hotter than He-WDs making a
direct comparison complicated, but the contrast between water
incidence rates for the two groups remains despite our addition
of several H-WDs over the work of \citet{Brouwers2023}.

As another approach, we apply a Kolmogorov-Smirnov (KS) test using the kstest function from \textsc{SciPy} \citep{2020SciPy-NMeth} to test whether the water mass fractions for the H and He sub-samples are statistically distinguishable from a null ``dry" distribution with a mean $\mathrm{f_{H_2O}}$ of 0. For a range of widths of the null distribution (from 0.05 to 0.5), and imposing a confidence level of $95\%$, the H-WD distribution is not distinguishable from the null distribution. The He-WD sample is distinguishable from the null, with a p-value $\ll0.01$, and is additionally distinguishable from the H sample with a p-value of $0.01$. The full sample (H- and He-WDs) is also consistent with a non-dry distribution. 

Finally, we also test the variation of p-value with the median of the null distribution, for the same range of null distribution widths from 0.05 to 0.5, and imposing a confidence level of $95\%$. The H sample remains indistinguishable from the null for $\mathrm{f_{H_2O}}\lessapprox45\%$, while the He sample is indistinguishable from the null for medians of $\approx15-45\%$. Therefore, while we rule out a ``dry" population of polluting bodies for the He-WDs, we cannot do so for the H sample. Additionally, we rule out an overall population of polluting bodies that are at least $50\%$ water by mass for both He and H WDs, though a relatively large range of median $\mathrm{f_{H_2O}}$ remains.

\startlongtable 
\tabletypesize{\small}
\begin{deluxetable*}{llllllll}
\tablecaption{Summary of water mass percentages ($\mathrm{f_{H_2O}}\times100$) derived through different methods. WDs are in the same order as Table \ref{table:WDsample}, grouped by atmospheric type (H above the line, He below) and listed in order of increasing $T_\mathrm{{eff}}$}.\label{tab:water_summary}
\tablehead{\colhead{WD} & \colhead{Buildup} & \colhead{Steady State} & \colhead{Settling} & \colhead{Original} & \colhead{Weighted} & \colhead{Restricted $\mathrm{\tau_d}$} & \colhead{\hspace{-3.5mm} Weighted +} \\*[-3mm]
\colhead{} & \colhead{} & \colhead{} & \colhead{} & \colhead{} & \colhead{} & \colhead{} & \colhead{Restricted $\mathrm{\tau_d}$ }}
\startdata
G29$-$38 & $5\pm_{5}^{16}$ & $0\pm_{0}^{8}$ & $0\pm_{0}^{7}$ & $0\pm_{0}^{12}$ & $0\pm_{0}^{9}$ & $0\pm_{0}^{7}$ & $0\pm_{0}^{2}$ \\
WD\,0145+234 & $9\pm_{9}^{23}$ & $19\pm_{19}^{22}$ & $19\pm_{19}^{22}$ & $5\pm_{5}^{32}$ & $9\pm_{9}^{32}$ & $10\pm_{10}^{28}$ & $8\pm_{8}^{37}$ \\
WD\,0842+572 & $0\pm_{0}^{0}$ & $0\pm_{0}^{7}$ & $0\pm_{0}^{7}$ & $0\pm_{0}^{0}$ & $0\pm_{0}^{4}$ & $0\pm_{0}^{2}$ & $0\pm_{0}^{7}$ \\
J0611$-$6931 & $27\pm_{16}^{14}$ & $39\pm_{14}^{12}$ & $39\pm_{14}^{12}$ & $0\pm_{0}^{46}$ & $0\pm_{0}^{23}$ & $26\pm_{20}^{16}$ & $29\pm_{17}^{14}$ \\
SDSS\,J1043+0855 & $0\pm_{0}^{9}$ & $0\pm_{0}^{27}$ & $0\pm_{0}^{27}$ & $0\pm_{0}^{0}$ & $0\pm_{0}^{31}$ & $0\pm_{0}^{16}$ & $0\pm_{0}^{20}$ \\
PG\,1015+161 & $0\pm_{0}^{0}$ & $0\pm_{0}^{0}$ & $0\pm_{0}^{0}$ & $0\pm_{0}^{0}$ & $0\pm_{0}^{0}$ & $0\pm_{0}^{0}$ & $0\pm_{0}^{0}$ \\
J0510+2315 & $14\pm_{13}^{13}$ & $37\pm_{13}^{12}$ & $37\pm_{13}^{12}$ & $82\pm_{37}^{18}$ & $35\pm_{12}^{12}$ & $38\pm_{13}^{12}$ & $21\pm_{11}^{11}$ \\
SDSS\,1228+1040 & $0\pm_{0}^{15}$ & $6\pm_{6}^{22}$ & $6\pm_{6}^{22}$ & $0\pm_{0}^{15}$ & $0\pm_{0}^{22}$ & $0\pm_{0}^{24}$ & $0\pm_{0}^{29}$ \\
GALEX\,1931+0117 & $0\pm_{0}^{4}$ & $0\pm_{0}^{19}$ & $0\pm_{0}^{19}$ & $0\pm_{0}^{0}$ & $0\pm_{0}^{0}$ & $0\pm_{0}^{8}$ & $0\pm_{0}^{8}$ \\
J0006+2858 & $2\pm_{2}^{11}$ & $25\pm_{11}^{11}$ & $25\pm_{11}^{11}$ & $15\pm_{15}^{17}$ & $23\pm_{12}^{11}$ & $21\pm_{15}^{13}$ & $11\pm_{11}^{9}$ \\
PG\,0843+516 & $0\pm_{0}^{0}$ & $0\pm_{0}^{0}$ & $0\pm_{0}^{0}$ & $0\pm_{0}^{0}$ & $0\pm_{0}^{0}$ & $0\pm_{0}^{0}$ & $0\pm_{0}^{0}$ \\
WD\,J0649$-$7624 & $0\pm_{0}^{5}$ & $4\pm_{4}^{24}$ & $4\pm_{4}^{24}$ & $0\pm_{0}^{25}$ & $1\pm_{1}^{24}$ & $3\pm_{3}^{28}$ & $0\pm_{0}^{12}$ \\
\hline
WD\,0446$-$255 & $15\pm_{13}^{12}$ & $14\pm_{13}^{12}$ & $3\pm_{3}^{10}$ & $3\pm_{3}^{21}$ & $2\pm_{2}^{18}$ & $5\pm_{5}^{16}$ & $4\pm_{4}^{21}$ \\
WD\,1350$-$162 & $17\pm_{12}^{11}$ & $16\pm_{12}^{11}$ & $3\pm_{3}^{9}$ & $0\pm_{0}^{2}$ & $0\pm_{0}^{3}$ & $0\pm_{0}^{4}$ & $0\pm_{0}^{4}$ \\
WD\,1232+563 & $51\pm_{14}^{12}$ & $50\pm_{14}^{12}$ & $37\pm_{14}^{13}$ & $35\pm_{18}^{15}$ & $35\pm_{15}^{15}$ & $36\pm_{14}^{14}$ & $36\pm_{22}^{11}$ \\
SDSS\,J1242+5226 & $48\pm_{11}^{10}$ & $48\pm_{11}^{10}$ & $39\pm_{11}^{10}$ & $32\pm_{18}^{24}$ & $33\pm_{23}^{24}$ & $35\pm_{23}^{20}$ & $36\pm_{23}^{20}$ \\
1013+0259 & $39\pm_{38}^{29}$ & $38\pm_{38}^{29}$ & $28\pm_{28}^{31}$ & $47\pm_{47}^{22}$ & $38\pm_{38}^{30}$ & $37\pm_{37}^{24}$ & $42\pm_{42}^{24}$ \\
SDSS\,J2339$-$0424 & $52\pm_{6}^{5}$ & $51\pm_{6}^{5}$ & $40\pm_{6}^{6}$ & $42\pm_{15}^{13}$ & $40\pm_{10}^{11}$ & $38\pm_{15}^{15}$ & $39\pm_{12}^{13}$ \\
SDSS\,J0738+1835 & $40\pm_{19}^{16}$ & $39\pm_{19}^{16}$ & $32\pm_{19}^{17}$ & $0\pm_{0}^{0}$ & $0\pm_{0}^{0}$ & $0\pm_{0}^{0}$ & $0\pm_{0}^{0}$ \\
HS\,2253+8023 & $22\pm_{7}^{7}$ & $21\pm_{7}^{7}$ & $5\pm_{5}^{6}$ & $11\pm_{11}^{9}$ & $11\pm_{10}^{10}$ & $11\pm_{9}^{10}$ & $9\pm_{9}^{10}$ \\
WD\,1425+540 & $71\pm_{18}^{12}$ & $71\pm_{18}^{12}$ & $57\pm_{21}^{16}$ & $58\pm_{30}^{21}$ & $64\pm_{25}^{16}$ & $61\pm_{32}^{20}$ & $64\pm_{24}^{20}$ \\
0944$-$0039 & $47\pm_{8}^{7}$ & $42\pm_{8}^{7}$ & $38\pm_{8}^{7}$ & $37\pm_{9}^{8}$ & $37\pm_{9}^{7}$ & $38\pm_{8}^{7}$ & $37\pm_{7}^{8}$ \\
Gaia\,J0218+3625 & $38\pm_{17}^{14}$ & $37\pm_{17}^{14}$ & $23\pm_{16}^{15}$ & $22\pm_{15}^{21}$ & $25\pm_{25}^{24}$ & $23\pm_{23}^{25}$ & $24\pm_{23}^{21}$ \\
EC\,22211$-$2525 & $26\pm_{17}^{15}$ & $26\pm_{17}^{15}$ & $9\pm_{9}^{14}$ & $15\pm_{15}^{25}$ & $13\pm_{13}^{20}$ & $6\pm_{6}^{28}$ & $14\pm_{14}^{22}$ \\
WD\,2207+121 & $27\pm_{17}^{15}$ & $26\pm_{17}^{15}$ & $14\pm_{14}^{15}$ & $16\pm_{16}^{16}$ & $15\pm_{15}^{20}$ & $16\pm_{16}^{19}$ & $13\pm_{13}^{19}$ \\
WD\,1551+175 & $23\pm_{16}^{15}$ & $22\pm_{16}^{15}$ & $8\pm_{8}^{15}$ & $16\pm_{16}^{18}$ & $7\pm_{7}^{17}$ & $13\pm_{13}^{15}$ & $10\pm_{10}^{15}$ \\
WD\,1244+498 & $34\pm_{15}^{14}$ & $32\pm_{15}^{14}$ & $11\pm_{11}^{14}$ & $28\pm_{26}^{20}$ & $22\pm_{22}^{19}$ & $20\pm_{20}^{28}$ & $28\pm_{27}^{18}$ \\
WD\,1248+1004 & $46\pm_{15}^{13}$ & $44\pm_{15}^{13}$ & $29\pm_{15}^{14}$ & $31\pm_{29}^{20}$ & $31\pm_{26}^{23}$ & $31\pm_{29}^{22}$ & $30\pm_{27}^{22}$ \\
GD\,40 & $7\pm_{7}^{15}$ & $6\pm_{6}^{14}$ & $0\pm_{0}^{4}$ & $0\pm_{0}^{12}$ & $0\pm_{0}^{19}$ & $0\pm_{0}^{15}$ & $0\pm_{0}^{16}$ \\
G241$-$6 & $23\pm_{14}^{13}$ & $22\pm_{14}^{13}$ & $10\pm_{10}^{12}$ & $3\pm_{3}^{19}$ & $5\pm_{5}^{21}$ & $0\pm_{0}^{19}$ & $6\pm_{6}^{17}$ \\
1516$-$0040 & $37\pm_{5}^{5}$ & $34\pm_{5}^{5}$ & $29\pm_{5}^{5}$ & $29\pm_{6}^{7}$ & $27\pm_{5}^{7}$ & $27\pm_{7}^{6}$ & $29\pm_{8}^{5}$ \\
Gaia\,J1922+4709 & $0\pm_{0}^{2}$ & $0\pm_{0}^{1}$ & $0\pm_{0}^{0}$ & $0\pm_{0}^{0}$ & $0\pm_{0}^{1}$ & $0\pm_{0}^{3}$ & $0\pm_{0}^{5}$ \\
WD\,1145+017 & $1\pm_{1}^{18}$ & $0\pm_{0}^{17}$ & $0\pm_{0}^{4}$ & $0\pm_{0}^{32}$ & $0\pm_{0}^{22}$ & $0\pm_{0}^{28}$ & $0\pm_{0}^{27}$ \\
GD\,378 & $71\pm_{6}^{6}$ & $70\pm_{6}^{6}$ & $55\pm_{8}^{7}$ & $63\pm_{9}^{7}$ & $64\pm_{10}^{7}$ & $64\pm_{7}^{7}$ & $63\pm_{11}^{7}$ \\
0859+1123 & $42\pm_{9}^{8}$ & $41\pm_{9}^{8}$ & $34\pm_{9}^{9}$ & $31\pm_{11}^{10}$ & $29\pm_{12}^{11}$ & $31\pm_{9}^{7}$ & $32\pm_{10}^{8}$ \\
0030+1526 & $54\pm_{8}^{7}$ & $53\pm_{8}^{7}$ & $42\pm_{9}^{9}$ & $46\pm_{9}^{8}$ & $48\pm_{10}^{6}$ & $45\pm_{9}^{11}$ & $48\pm_{11}^{7}$ \\
0930+0618 & $62\pm_{4}^{4}$ & $61\pm_{4}^{4}$ & $53\pm_{5}^{4}$ & $54\pm_{6}^{5}$ & $53\pm_{5}^{6}$ & $54\pm_{6}^{5}$ & $53\pm_{6}^{5}$ \\
1627+1723 & $17\pm_{17}^{28}$ & $15\pm_{15}^{28}$ & $7\pm_{7}^{28}$ & $5\pm_{5}^{31}$ & $14\pm_{14}^{34}$ & $8\pm_{8}^{38}$ & $4\pm_{4}^{35}$ \\
1109+1318 & $52\pm_{13}^{11}$ & $50\pm_{14}^{11}$ & $42\pm_{14}^{12}$ & $46\pm_{15}^{15}$ & $48\pm_{13}^{13}$ & $46\pm_{16}^{14}$ & $42\pm_{14}^{14}$ \\
SDSS\,J1734+6052 & $23\pm_{17}^{16}$ & $21\pm_{16}^{16}$ & $6\pm_{6}^{15}$ & $4\pm_{4}^{27}$ & $10\pm_{10}^{23}$ & $0\pm_{0}^{28}$ & $7\pm_{7}^{22}$ \\
0259$-$0721 & $38\pm_{5}^{5}$ & $36\pm_{5}^{5}$ & $30\pm_{5}^{5}$ & $0\pm_{0}^{0}$ & $0\pm_{0}^{0}$ & $0\pm_{0}^{0}$ & $0\pm_{0}^{0}$ \\
GD\,424 & $6\pm_{6}^{12}$ & $4\pm_{4}^{12}$ & $0\pm_{0}^{9}$ & $0\pm_{0}^{7}$ & $0\pm_{0}^{9}$ & $0\pm_{0}^{9}$ & $0\pm_{0}^{9}$ \\
1359$-$0217 & $53\pm_{11}^{10}$ & $51\pm_{11}^{10}$ & $43\pm_{12}^{11}$ & $46\pm_{19}^{11}$ & $43\pm_{13}^{11}$ & $46\pm_{15}^{13}$ & $44\pm_{13}^{11}$ \\
J0644$-$0352 & $16\pm_{16}^{15}$ & $14\pm_{14}^{15}$ & $1\pm_{1}^{15}$ & $3\pm_{3}^{22}$ & $0\pm_{0}^{26}$ & $0\pm_{0}^{19}$ & $3\pm_{3}^{16}$ \\
GD\,61 & $32\pm_{16}^{14}$ & $29\pm_{16}^{15}$ & $23\pm_{16}^{15}$ & $13\pm_{13}^{27}$ & $14\pm_{14}^{24}$ & $16\pm_{16}^{23}$ & $11\pm_{11}^{25}$ \\
WD\,1415+234 & $0\pm_{0}^{0}$ & $0\pm_{0}^{0}$ & $0\pm_{0}^{0}$ & $0\pm_{0}^{0}$ & $0\pm_{0}^{0}$ & $0\pm_{0}^{0}$ & $0\pm_{0}^{0}$ \\
SDSS\,J2248+2632 & $16\pm_{16}^{17}$ & $13\pm_{13}^{17}$ & $1\pm_{1}^{16}$ & $0\pm_{0}^{25}$ & $0\pm_{0}^{16}$ & $0\pm_{0}^{20}$ & $0\pm_{0}^{23}$ \\
WD\,J2047$-$1259 & $30\pm_{13}^{12}$ & $27\pm_{13}^{12}$ & $19\pm_{13}^{14}$ & $13\pm_{13}^{20}$ & $20\pm_{18}^{13}$ & $18\pm_{18}^{13}$ & $15\pm_{15}^{20}$ \\
Ton\,345 & $0\pm_{0}^{1}$ & $0\pm_{0}^{0}$ & $0\pm_{0}^{0}$ & $0\pm_{0}^{0}$ & $0\pm_{0}^{0}$ & $0\pm_{0}^{0}$ & $0\pm_{0}^{0}$ \\
WD\,1536+520 & $22\pm_{17}^{17}$ & $13\pm_{13}^{17}$ & $11\pm_{11}^{17}$ & $5\pm_{5}^{19}$ & $3\pm_{3}^{27}$ & $0\pm_{0}^{29}$ & $4\pm_{4}^{21}$ \\
WD\,1622+587 & $0\pm_{0}^{0}$ & $0\pm_{0}^{0}$ & $0\pm_{0}^{0}$ & $0\pm_{0}^{0}$ & $0\pm_{0}^{0}$ & $0\pm_{0}^{0}$ & $0\pm_{0}^{0}$ \\
\enddata
\tablecomments{The water calculation methods are as follows. 
``Buildup", ``Steady State", ``Settling": Observed abundances are adjusted assuming the WD is observed in the mass-buildup phase at time $t_{\mathrm{buildup}} = 0.5 \times \min ( \tau_{d}, \tau_{z})$, in steady state at the median $t_{\mathrm{SS}}(\mathrm{Z})$ (Equation \ref{eq:J09_tSS}), and in the mass-settling phase at time $t_{\mathrm{settling}} = 2 \times \max ( \tau_{d}, \tau_{z})$, respectively (Section \ref{section:picking_phases}). $\tau_{d}$ and $\tau_{z}$ are the disk lifetime and element settling times, respectively. ``Original", ``Weighted", ``Restricted $\mathrm{\tau_d}$" and ``Weighted + Restricted $\mathrm{\tau_d}$" refer to the four MCMC tests with different likelihood functions, calculated on the assumption that initial parent body compositions are chondritic. ``Original" allowed the parameters to vary freely, while the ``Weighted" favors solutions that maximize mass in the WD atmosphere, and the ``Restricted" tests placed a more conservative prior on the disk lifetime (see Sections \ref{section:mcmc_methods} and \ref{section:mcmc_solutions} for more details). }
\end{deluxetable*}

We do not consider it likely that H- and He-WDs are sampling different planetary system material as this would imply a planetary system somehow
having knowledge of what type of stellar remnant its host star will
turn into. Barring the identification of such a mechanism, 
if H- and He-WDs continue to disagree on average as to
water incidence rates $-$ especially as more H-WDs are added to the sample
across all temperatures $-$ then it will be necessary to carefully
assess all assumptions about how we interpret atmospheric pollution
for WDs.

\section{Conclusions}\label{section:conclusion}
We present new observations for three H-WDs
where two show potential evidence for having accreted water-rich
parent bodies. Combining these results with a large sample of 
previously published polluted WD data we infer water mass fractions for exoplanetary material based on excesses of oxygen relative to other rock-forming elements. We calculate minimum water mass fractions for each sample by assuming the accreted body had no significant metal component. 

To explore the impact of different accretion phases and assumed disk lifetimes on the elemental abundance ratios and resulting water abundance for WD pollution we assess accretion in the mass-buildup, steady state, and mass-settling phases.
We show that inferred abundance ratios and water fractions are usually within the uncertainty limits of the observed values through the mass-settling phase if the disk lifetime is longer than the settling timescales (as for most H-WDs), or until late times in the mass-settling phase if the disk lifetime is less than the settling timescales (most He-WDs). For a simple exponential accretion and settling model, correcting abundances for an assumed phase of accretion therefore has a minimal effect except for He-WDs observed after a few settling timescales (assuming typical disk lifetimes of $10^5-10^6$ yr), or unless true disk lifetimes are much shorter than current estimates. 

Under the hypothesis that WD pollution is typically chondritic in composition $-$ supported by the pollution compositions in this work $-$ 
we ran several MCMC tests to outline the distribution of accretion parameters that best explain the observed pollution. From these tests, we find that 
H-WDs are consistent with being in the steady state phase.
For He-WDs we generally find very broad posterior distributions for the disk lifetime and observation times, a result of abundance ratios being relatively stable except at the very early or very late stages of accretion.
While He-WDs are not generally well-constrained, they are compatible with
being in the mass-buildup phase. There are some uncertainties that we do not account for (see Section \ref{sec:3.1}) which may affect the reported water mass fractions. 

Overall, the median water content for the polluters in our sample is about $25\%$ by mass, consistent with estimates for typical icy solar system bodies. Uncertainties in water concentrations, propagated from uncertainties in measured elemental abundances, are in the range of $5-30\%$, with a median of $15\%$. Based on the results of our MCMC (the ``Original" case), 31/51 WDs have posterior distributions where $\geq50\%$ of the draws result in non-zero water abundances. The ``Weighted", ``Restricted" and ``Weighted + Restricted" cases similarly have 31, 29, and 32 WDs with non-zero median water concentrations. 
For the He-WDs, 34/39 of the inferred water abundances are supported by sufficient quantities of H in the WD atmosphere; further analysis should be done on the remaining 5 WDs to determine whether the relative lack of H is physical. We are not able to conclusively identify post-main sequence evaporation and/or asynchronous accretion as having an impact on translating the observed water mass fractions to main sequence parent body properties. 
Within the context of this analysis, the conclusion that the polluting material originates from a largely water-rich (median $\mathrm{f_{H_2O}}\approx25\%$) population is also supported by applying a KS test to the distribution of water mass fractions. For the He and combined He and H samples, we can reject the null hypothesis of a dry distribution of pollution parent bodies. 

Finally, there is a significant difference in water mass fractions for pollution in H versus He atmospheres. 
We find that pollution in H-WDs tends to lack significant O excesses compared to He-WDs. 
While pollution in He atmospheres is consistent with an overall water-rich population of bodies, pollution in H atmospheres tends to be more dry.  
It is unlikely that H and He WDs sample different populations of material, and an increased sample of polluted H-WDs at lower temperatures would help untangle this apparent mismatch.

\acknowledgements
We thank B. Zuckerman and B. Klein for their contributions to this work. 
ILT was supported by the Future Investigators in NASA Earth and Space Science and Technology (FINESST) grant 80NSSC23K1383. EDY and ILT were also supported by the NASA Exobiology grant 80NSSC20K0270. 
CM acknowledges support from NSF grants SPG-1826583 and SPG-1826550.
S. Xu is supported by the international Gemini Observatory, a program of NSF NOIRLab, which is managed by the Association of Universities for Research in Astronomy (AURA) under a cooperative agreement with the U.S. National Science Foundation, on behalf of the Gemini partnership of Argentina, Brazil, Canada, Chile, the Republic of Korea, and the United States of America.
This project has received funding from the European Research Council (ERC) under the European Union’s Horizon 2020 research and innovation programme (Grant agreement No. 101020057).
This paper is based on observations made with the NASA/ESA $Hubble$ $Space$ $Telescope$, 
obtained at the Space Telescope Science Institute, which is operated by the
Association of Universities for Research in Astronomy, Inc., under NASA contract NAS 5-26555.
This work was supported by associated grants HST-GO-15817.001-A
and HST-GO-16032.001-A. This paper includes data gathered with the 6.5 meter Magellan Telescopes located at Las Campanas Observatory, Chile.
Some of the data presented herein were obtained at the W.M. Keck Observatory, 
which is operated as a scientific partnership among the California Institute of 
Technology, the University of California and the National Aeronautics and Space 
Administration. The Observatory was made possible by the generous financial 
support of the W.M. Keck Foundation.
The authors wish to recognize and acknowledge the very significant cultural role and 
reverence that the summit of Mauna Kea has always had within the indigenous Hawaiian 
community.  We are most fortunate to have the opportunity to conduct observations from 
this mountain.

\appendix

\section{Photospheric Spectral Lines for H-WDs}\label{appsection:line_lists}

Tables \ref{tab:0842_lines}, \ref{tab:0145_lines}, and \ref{tab:0649_lines} list the details of the spectral line measurements for the three H-dominated white dwarfs WD\,0842+572, WD\,0145+234, and WD\,J0649$-$7624, respectively. We measure equivalent widths (EWs) using the IRAF \textit{splot} task, and calculate resulting radial velocities ($\mathrm{V_r}$) from the Doppler shifts of the identified line centers from laboratory wavelengths. Equivalent width uncertainties are typically $\sim10\%$, and optical and UV radial velocity uncertainties are around $3~\rm km/s$ and $5-10~\rm km/s$, respectively. 

\tabletypesize{\scriptsize}
\begin{deluxetable}{llrr}
\tablecaption{Photospheric absorption lines for WD\,0842+572} \label{tab:0842_lines} 
\tablehead{ \colhead{Ion} & \colhead{$\lambda$ (\r{A})} & \colhead{EW (m\r{A})} & \colhead{$\mathrm{V_r}$} }
\startdata
O I & 7771.94 & 57 & 24 \\
O I & 7774.16 & 51 & 25 \\
O I & 7775.39 & 59 & 23 \\
Mg I & 3829.35 & 14 & 22 \\
Mg I & 3832.30 & 32 & 24 \\
Mg I & 3838.29 & 50 & 23 \\
Mg I & 5172.68 & 39 & 27 \\
Mg I & 5183.60 & 58 & 26 \\
Mg II & 4481.33 & 813 & 23 \\ 
Mg II & 7877.05 & 104 & 27 \\
Mg II & 7896.36 & 175 & 22 \\
Si II & 3856.02 & 57 & 25 \\
Si II & 4128.05 & 32 & 25 \\
Si II & 4130.89 & 64 & 24 \\
Si II & 5055.98 & 154 & 31 \\
Si II & 6347.11 & 167 & 22 \\
Si II & 6371.37 & 86 & 22 \\
Ca II & 3158.87 & 61 & 27 \\
Ca II & 3179.33 & 80 & 29 \\
Ca II & 3181.28 & 34 & 28 \\
Ca II & 3706.02 & 13 & 27 \\
Ca II & 3933.66 & 140 & 24 \\
Ca II & 3968.47 & 29 & 26 \\
Ti II & 3234.51 & 8 & 25 \\
Ti II & 3341.87 & 12 & 26 \\
Ti II & 3349.03 & 21 & 29 \\
Ti II & 3349.40 & 22 & 27 \\
Ti II & 3372.79 & 12 & 26 \\
Ti II & 3383.76 & 6 & 25 \\
Cr II & 3132.05 & 24 & 27 \\
Cr II & 3368.04 & 20 & 25 \\
Mn II & 3441.99 & 16 & 25 \\
Fe I & 3581.19 & 9 & 27 \\
Fe II & 3154.20 & 33 & 24 \\
Fe II & 3167.86 & 25 & 25 \\
Fe II & 3177.53 & 32 & 28 \\
Fe II & 3186.74 & 19 & 25 \\
Fe II & 3193.86 & 24 & 21 \\
Fe II & 3196.07 & 16 & 26 \\
Fe II & 3210.44 & 19 & 26 \\
Fe II & 3213.31 & 50 & 24 \\
Fe II & 3227.74 & 40 & 25 \\
Fe II & 3259.05 & 10 & 26 \\
Fe II & 4233.17 & 16 & 23 \\
Fe II & 4522.63 & 9 & 23 \\
Fe II & 4583.84 & 14 & 24 \\
Fe II & 4583.84 & 15 & 24 \\
Fe II & 5234.62 & 11 & 25 \\
Fe II & 5316.61 & 14 & 24 \\
Ni II & 3513.99 & 11 & 23 \\
\enddata
\end{deluxetable}

\tabletypesize{\scriptsize}
\begin{deluxetable}{llrr}
\tablecaption{Photospheric absorption lines for WD\,0145+234}\label{tab:0145_lines}
\tablehead{ \colhead{Ion} & \colhead{$\lambda$ (\r{A})} & \colhead{EW (m\r{A})} & \colhead{$\mathrm{V_r}$} }
\startdata
O I & 1152.15 & 30 & 34 \\
O I & 1304.86 & 74 & 51 \\
O I & 1306.03 & 49 & 53 \\
Mg I & 5172.68 & 8 & 49 \\
Mg I & 5183.60 & 11 & 47 \\
Mg II & 4481.33 & 40 & 41 \\ 
Al II & 1670.79 & 304 & 40 \\
Al II & 1719.44 & 30 & 51 \\
Al II & 1721.27 & 66 & 37 \\
Al II & 1724.98 & 106 & 41 \\
Al II & 1760.10 & 31 & 54 \\ 
Al II & 1763.95 & 78 & 44 \\
Si II & 1264.74 & 257 & 35 \\
Si II & 1309.28 & 107 & 50 \\ 
Si II  & 1348.54 & 23 & 49 \\
Si II  & 1350.07 & 24 & 50 \\
Si II & 1352.64 & 33 & 33 \\ 
Si II  & 1353.72 & 42 & 35 \\ 
Si II & 1533.43 & 237 & 37 \\
Ca II & 3158.87 & 45 & 47 \\
Ca II & 3179.33 & 61 & 45 \\
Ca II & 3706.02 & 6 & 48 \\
Ca II & 3736.90 & 20 & 43 \\
Ca II & 3933.66 & 156 & 44 \\
Ca II & 3968.47 & 42 & 44 \\
Ca II & 8498.02 & 22 & 42 \\
Ca II & 8542.09 & 92 & 43 \\
Ca II & 8662.14 & 69 & 44 \\
Ti II & 3234.51 & 10 & 44 \\
Ti II & 3349.40 & 10 & 45 \\
Ti II & 3361.21 & 8 & 42 \\
Ti II & 3372.79 & 8 & 43 \\
Fe I & 3749.49 & 5 & 46 \\
Fe II & 5018.44 & 2 & 45 \\ 
Fe II & 5169.03 & 4 & 45 \\ 
Fe II  & 1311.06 & 40 & 31 \\ 
Fe II  & 1383.58 & 28 & 56 \\ 
Fe II  & 1412.84 & 25 & 49 \\
Fe II  & 1424.78 & 25 & 41 \\
Ni II & 1411.07 & 23 & 49 \\
\enddata
\end{deluxetable}

\tabletypesize{\scriptsize}
\begin{deluxetable}{llrr}
\tablecaption{Photospheric absorption lines for WD\,J0649$-$7624}\label{tab:0649_lines}
\tablehead{ \colhead{Ion} & \colhead{$\lambda$ (\r{A})} & \colhead{EW (m\r{A})} & \colhead{$\mathrm{V_r}$} }
\startdata
O I & 1152.15 & 40 & 24 \\
O I & 1304.86 & 40 & 29 \\
Mg II & 4481.33 & 82 & 26 \\ 
Si II  & 1194.50 & 69 & 27 \\
Si II  & 1197.39 & 28 & 30 \\
Si II  & 1246.74 & 21 & 24 \\
Si II  & 1248.43 & 26 & 29 \\
Si II  & 1250.09 & 20 & 32 \\
Si II  & 1251.16 & 18 & 31 \\
Si II  & 1264.73 & 339 & 33 \\ %
Si II  & 1305.59 & 45 & 37 \\ 
Si II  & 1309.45 & 147 & 35 \\ 
Si II  & 1346.88 & 23 & 32 \\
Si II  & 1348.54 & 28 & 29 \\
Si II  & 1350.07 & 38 & 26 \\
Si II  & 1352.64 & 35 & 24 \\
Si II  & 6347.11 & 65 & 25 \\
Si III & 1301.15 & 106 & 38 \\
Si III & 1303.32 & 86 & 28 \\
Si III & 1312.59 & 34 & 36 \\
Si IV & 1393.78 & 116 & 24 \\ %
Si IV & 1402.77 & 86 & 30 \\
\enddata
\end{deluxetable}

\bibliography{bibli}{}
\bibliographystyle{aasjournal}

\end{CJK}
\end{document}